\documentclass[a4paper,twoside]{article}

\newcommand{\affil}[1]{$^{\rm #1}$}
\usepackage[authoryear]{natbib}
\bibpunct{(}{)}{;}{a}{}{,}
\usepackage{graphicx}

\date{} 
\title{\large\bf\flushleft The Murchison Widefield Array: the Square Kilometre Array Precursor at low radio frequencies}
\author{\parbox{\textwidth}{\flushleft
\vspace{-0.5cm}
{\it S.J. Tingay\affil{A,S,T}, R. Goeke\affil{B}, J.D. Bowman\affil{C}, D. Emrich\affil{A}, S.M. Ord\affil{A}, D.A. Mitchell\affil{D,S}, M.F. Morales\affil{E}, T. Booler\affil{A}, B. Crosse\affil{A}, D. Pallot\affil{A}, A. Wicenec\affil{F}, W. Arcus\affil{A}, D. Barnes\affil{G}, G. Bernardi\affil{H}, F. Briggs\affil{I,S}, S. Burns\affil{J}, J.D. Bunton\affil{K}, R.J. Cappallo\affil{L}, T. Colegate\affil{A}, B.E. Corey\affil{L}, A. Deshpande\affil{M}, L. deSouza\affil{K}, B.M. Gaensler\affil{N,S}, L.J. Greenhill\affil{H}, P.J. Hall\affil{A}, B.J. Hazelton\affil{E}, D. Herne\affil{A}, J.N. Hewitt\affil{B}, M. Johnston-Hollitt\affil{O}, D.L. Kaplan\affil{P}, J.C. Kasper\affil{H}, B.B. Kincaid\affil{L}, R. Koenig\affil{K}, E. Kratzenberg\affil{L}, C.J. Lonsdale\affil{L}, M.J. Lynch\affil{A}, B. McKinley\affil{I,S}, S.R. McWhirter\affil{L}, E. Morgan\affil{B}, D. Oberoi\affil{Q}, J. Pathikulangara\affil{K}, T. Prabu\affil{M}, R.A Remillard\affil{B}, A.E.E. Rogers\affil{L}, A. Roshi\affil{R}, J.E. Salah\affil{L}, R.J. Sault\affil{D}, N. Udaya-Shankar\affil{M}, F. Schlagenhaufer\affil{A}, K.S. Srivani\affil{M}, J. Stevens\affil{K}, R. Subrahmanyan\affil{M,S}, S. Tremblay\affil{A,S}, R.B. Wayth\affil{A,S}, M. Waterson\affil{A}, R.L. Webster\affil{D,S}, A.R. Whitney\affil{L}, A. Williams\affil{F}, C.L. Williams\affil{B} and J.S.B. Wyithe\affil{D,S}}\\
\vspace{0.4cm}
{\small \affil{A}\,ICRAR - Curtin University, Perth, Australia}\\
{\small \affil{B}\,MIT Kavli Institute for Astrophysics and Space Research, Cambridge, MA, USA}\\
{\small \affil{C}\,Arizona State University, Tempe, AZ, USA}\\
{\small \affil{D}\,The University of Melbourne, Melbourne, Australia}\\
{\small \affil{E}\,University of Washington, Seattle, USA}\\
{\small \affil{F}\,ICRAR - University of Western Australia, Perth, Australia}\\
{\small \affil{G}\,Monash University, Melbourne, Australia}\\
{\small \affil{H}\,Harvard-Smithsonian Center for Astrophysics, Cambridge, MA, USA}\\
{\small \affil{I}\,The Australian National University, Canberra, Australia}\\
{\small \affil{J}\,Burns Industries, Nashua, NH, USA}\\
{\small \affil{K}\,CSIRO Astronomy and Space Science, Australia}\\
{\small \affil{L}\,MIT Haystack Observatory, Westford, MA, USA}\\
{\small \affil{M}\,Raman Research Institute, Bangalore, India}\\
{\small \affil{N}\,Sydney Institute for Astronomy, The University of Sydney, Sydney, Australia}\\
{\small \affil{O}\,School of Chemical and Physical Sciences, Victoria University of Wellington, New Zealand}\\
{\small \affil{P}\,University of Wisconsin--Milwaukee, Milwaukee, WI, USA}\\
{\small \affil{Q}\,National Centre for Radio Astrophysics, Pune, India}\\
{\small \affil{R}\,National Radio Astronomy Observatory, Charlottesville, WV, USA}\\
{\small \affil{S}\,ARC Centre of Excellence for All-sky Astrophysics (CAASTRO)}\\
{\small \affil{T}\,Email: s.tingay@curtin.edu.au}}}

\begin{document}
\twocolumn[
\begin{changemargin}{.8cm}{.5cm}
\begin{minipage}{.9\textwidth}
\vspace{-1cm}
\maketitle
%
%
\small{\bf Abstract:}
The Murchison Widefield Array (MWA) is one of three Square Kilometre Array Precursor telescopes and is located at the Murchison Radio-astronomy Observatory in the Murchison Shire of the mid-west of Western Australia, a location chosen for its extremely low levels of radio frequency interference.  The MWA operates at low radio frequencies, 80$-$300 MHz, with a processed bandwidth of 30.72 MHz for both linear polarisations, and consists of 128 aperture arrays (known as tiles) distributed over a $\sim$3 km diameter area.  Novel hybrid hardware/software correlation and a real-time imaging and calibration systems comprise the MWA signal processing backend.  In this paper the as-built MWA is described both at a system and sub-system level, the expected performance of the array is presented, and the science goals of the instrument are summarised.

\medskip
\medskip
\end{minipage}
\end{changemargin}
]
\small

\section{Introduction}
The Murchison Widefield Array (MWA\footnote{http://www.mwatelescope.org; http://www.facebook.com/Murchison.Widefield.Array}) is the Square Kilometre Array (SKA: \citet{ska}) Precursor telescope at low radio frequencies.  An SKA Precursor is a recognised SKA technology demonstrator located at one of the two sites shortlisted for the SKA in 2006, the Murchison Radio-astronomy Observatory (MRO) in the Murchison region of Western Australia and the Karoo region of South Africa's Northern Cape.  The MWA is one of two SKA Precursor telescopes sited at the MRO, the other being the Australian SKA Pathfinder (ASKAP: \citet{jon08}; \citet{jon07}).  The MeerKAT\footnote{http://public.ska.ac.za/meerkat} SKA Precursor is located at the South African site.  The MRO was chosen for the site of the MWA due to its extremely low levels of human-made radio frequency intereference, particularly in the FM band encompassed by the MWA at the low end of its operating frequency range \citep{bow10}.  The MRO has been chosen as the host site for the low frequency component of the SKA, in both Phases 1 and 2, to consist of sparse aperture arrays\footnote{http://www.skatelescope.org/news/dual-site-agreed-square-kilometre-array-telescope/}.

While only three instruments have SKA Precursor status, a number of other instruments have SKA Pathfinder status (SKA technology demonstrator but not on a candidate SKA site) and are also feeding information into the SKA design and costing process.  The most notable of the SKA Pathfinders in the MWA frequency range is LOFAR, built in The Netherlands \citep{lofar}.

The MWA and ASKAP are complementary, in that they operate in different frequency ranges (MWA: 80 - 300 MHz; ASKAP: 0.7 - 1.8 GHz) and employ different antenna technologies (MWA: aperture arrays (tiles); ASKAP: dishes plus Phased Array Feeds).  The combined MWA and ASKAP technology specifications almost fully sample the roadmap technologies for the SKA \citep{ska} at a single radio quiet location, the MRO.  In particular, the MWA explores the so-called large-N and small-D array concept that will be utilised for the SKA, with large numbers of small receiving elements providing a large field of view on the sky and therefore high sensitivity over wide fields, equating to high survey speed, a key metric for SKA science \citep{skascience}.

All three of the MWA, ASKAP and MeerKAT are planning, or have operated, demonstrator instruments.  ASKAP is building the six antenna Boolardy Engineering Test Array (BETA\footnote{http://www.atnf.csiro.au/projects/askap}).  MeerKAT is being preceeded by KAT7\footnote{http://public.ska.ac.za/kat-7}, a seven antenna array.  For the MWA, a 32 tile test array operated between 2009 and 2011, allowing the assesment of a number of generations of prototype hardware as well as several iterations of MWA system integration.  The MWA 32 tile array also undertook science-quality astronomical observations, in order to demonstrate the performance of the hardware on-sky.  A number of science results have been reported from the 32 tile data \citep{Williams-etal.2012, Bernardi-etal.2012, mck12, bel12, obe11}.  Operation of the 32 tile test array ceased in late 2011, in preparation for the commencement of construction for the final MWA instrument in early 2012, due for completion in late 2012.

Previously, \citet{Lonsdale-etal.2009} described the conceptual underpinings of the MWA design, the advantages of a large-N, small-D architecture for radio interferometers and some of the challenges inherent in the data processing for this type of telescope.  \citet{Lonsdale-etal.2009} also provided brief descriptions of the initial plans for MWA hardware and data processing elements, as well as the science goals for the MWA.

Since the \citet{Lonsdale-etal.2009} paper was published, a number of design and construction considerations, that were uncertain at the time of publication, have been finalised.  In particular the originally envisaged 512 tiles was re-scoped to 128 tiles due to funding constraints.  The construction of the final MWA instrument is underway and practical completion of the facility is expected in late 2012.  MWA commissioning will commence in mid-2012, with science operations to commence in approximately mid-2013.  Thus, the purpose of this paper is to provide a full description of the MWA in its as-built form, including the final system architecture and sub-systems, as well as signal processing and data handling strategies (Section 2).  This paper will also give a brief summary of the MWA science goals, a full description of which will appear elsewhere \citep{bow12} and details on the expected performance of the MWA (Sections 3 and 4).  This paper is intended to inform future users of the capabilities of the MWA, ahead of the science operations phase, such that users can commence planning for MWA science activities.  Section 5 briefly discusses the MWA within the context of other existing and future instrumentation, including the Phase 1 SKA.

\section{MWA system design and sub-system descriptions}
\begin{figure*}[ht]
\begin{center}
\includegraphics[scale=0.5, angle=0]{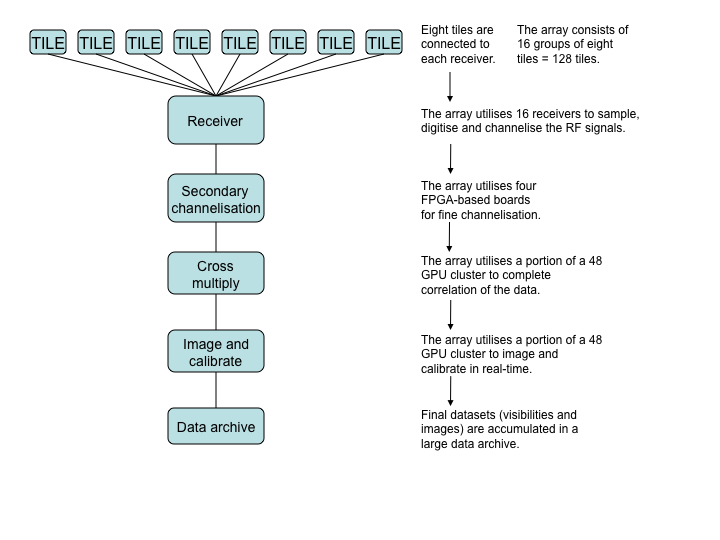}
\caption{High level schematic overview of MWA physical system and signal path}\label{fig:signalpath}
\end{center}
\end{figure*}

We start with a very brief overview of the MWA signal path, and provide details on each of the identified sub-systems in subsequent Sections.  Table \ref{tab:system} provides a summary of the MWA system parameters and expected performance.  Figure \ref{fig:signalpath} gives a high-level schematic overview of the MWA physical system and signal path.

\begin{table*}[ht]
\begin{center}
\caption{System parameters for MWA}\label{t_system}\label{tab:system}
\begin{tabular}{lccc}
\hline 
Parameter & Symbol & 150 MHz & 200 MHz \\
\hline
Number of tiles 				&	N 				& 	128 		&	128 \\
Area of one tile at zenith (m$^2$) 	& 	A$_{\rm eff}$ 	& 	21.5		&	19.8 \\
Total collecting area (m$^2$) 	&					& 	2752		&	2534 \\
Receiver temperature (K)		&	T$_{\rm rcv}$	& 	50 			&	25 \\
$^a$Typical sky temperature (K)	&	T$_{\rm sky}$	&	350		&	170 \\
$^{b}$Field of view (deg$^2$)		&	$\Omega_{\rm P}$	&	610		& 	375 \\
Instantaneous bandwidth (MHz)	&	B 			&	30.72			& 	30.72 \\
Spectral resolution (MHz)		&					&	0.04		& 	0.04 \\
Temporal Resolution				&			& 0.5 s uncalibrated	& 0.5 s uncalibrated \\
								&			& 8 s calibrated		& 8 s calibrated \\
Polarization					&					& Full Stokes	& Full Stokes \\
Minimum baseline (m)			&					&	7.7			&	7.7 \\
Maximum baseline (m)		&					&	2864		&	2864 \\
Angular resolution (1.5 km array)&&$\sim$3'&$\sim$2' \\
Angular resolution (3 km array)&&$\sim$2'&$\sim$1' \\
\hline
\end{tabular}
\medskip\\
$^a$\citet{ska113}. \\
$^{b}$ Based on FWHM of primary beam.  Imageable area is significantly larger. \\
\end{center}
\end{table*}

The MWA signal path starts with a dual-polarisation dipole antenna, roughly a square meter of collecting area at $\sim$150 MHz.  Sixteen of these antennas are configured as an aperture array on a regular 4x4 grid (with a spacing of 1.1 m). Their signals are combined in an analog beamformer, using a set of switchable delay lines to provide coarse pointing capability. Each beamformer produces two wideband analog outputs representing orthogonal X and Y linear polarisations.  This we refer to as an antenna tile and analog beamformer (Section 2.3).

The sum of our four science drivers (Section 4 and \citet{bow12}) leads to a desire for a $uv$ baseline distribution which has a dense core surrounded by a smooth $r^{-2}$ radial distribution (Section 2.2).  Our core area has 50 antenna tiles uniformly distributed over a 100 metre diameter core, surrounded by 62 tiles which are distributed over a 1.5 km diameter circle.   The final 16 tiles have been placed even further out on a 3 km diameter circle to optimise solar imaging performance, and for the highest angular resolution imaging.

A host of practical considerations led us to combine the signals from 8 tiles into a single receiver (Section 2.4); we thus deploy 16 receivers distributed over our landscape such that no tile is more than 500 metres removed from its associated receiver.  A receiver filters the two analog signals from each tile to a bandpass of 80 to 300 MHz, Nyquist samples the signals with an 8-bit A/D converter, and digitally filters the result into 24$\times$1.28 MHz frequency channels which form a selectable, usually but not necessarily contiguous, 30.72 MHz sample space.  The receivers themselves are a mixture of high technology and low, combining high-speed, and hence high power, digital circuitry with mechanical cooling.  They are powered from a standard 240 V mains circuit and send out their digital data streams on three 2.1 Gbps fiber optic links.  One of our early design decisions was to provide spare power and fibre connections to most receiver locations so that future instrument development could share our existing infrastructure at low incremental cost.

About 5 km from the core of our array sits a building provided by CSIRO, which we share with the ASKAP project (the Central Processing Facility: CPF).  It has power, Electrommagnetic Interference (EMI) shielding, and water-cooled equipment racks for our data processing hardware.  The data streams from all 16 receivers meet here and each of the 1.28~MHz channels is filtered by dedicated hardware into 128$\times$10 kHz fine channels (Section 2.5).  A correlator implemented in software using general purpose graphical processing units (GPGPUs) processes, averages in time and frequency space, and outputs its results in 768$\times$40 kHz channels with 0.5 s resolution (Section 2.5).  This 2.25 Gbps stream of correlation products is then processed by the MWA Real Time System (RTS: \citet{Mitchell-etal.2012}) to produce real-time calibrated images every 8 s (Section 2.6).  We have enabled the ability for visibility data to also be saved, to allow for the possibility of post-observation processing (Sections 2.6 and 2.7).

The MWA runs a monitor and control system to schedule observations and monitor system health and parameters of use for data processing (Section 2.8).

The RTS output and the $uv$ data are transmitted on a dedicated 10 Gbps optical fibre connection to the Pawsey High Performance Computing Centre for SKA science in Perth, where 15 PB of data storage capacity has been reserved over a 5 year period to accommodate the MWA data archive (Section 2.7).

Supporting the MWA instrument is the underlying infrastructure, including the reticulation of fibre and power around the array and the connection to MRO site-wide services provided by the Commonwealth Scientific and Industrial Research Organisation (CSIRO) (Section 2.1).

In the following Sections we provide further detail on the MWA sub-systems.
	
\subsection{MWA infrastructure at the MRO}

The MWA support infrastructure at the MRO includes all of the installations and equipment required to transport power and data to, from, and around the array. The focus of the MWA infrastructure is a power and data distribution ``hub" located near the core of the array. This Section describes the distribution hub and then, in turn, its connections to central MRO site services; and its connections to MWA equipment in the field. All of the infrastructure and equipment described in this paper conforms to the specification ``RFI Standards for Equipment to be Deployed on the MRO"\footnote{ASKAP-MRO-0001-Version1.1, dated 15 Oct 2010}.

\begin{figure*}[ht]
\begin{center}$
\begin{array}{cc}
\includegraphics[clip=true, trim=25 100 25 100, scale=0.5, angle=0]{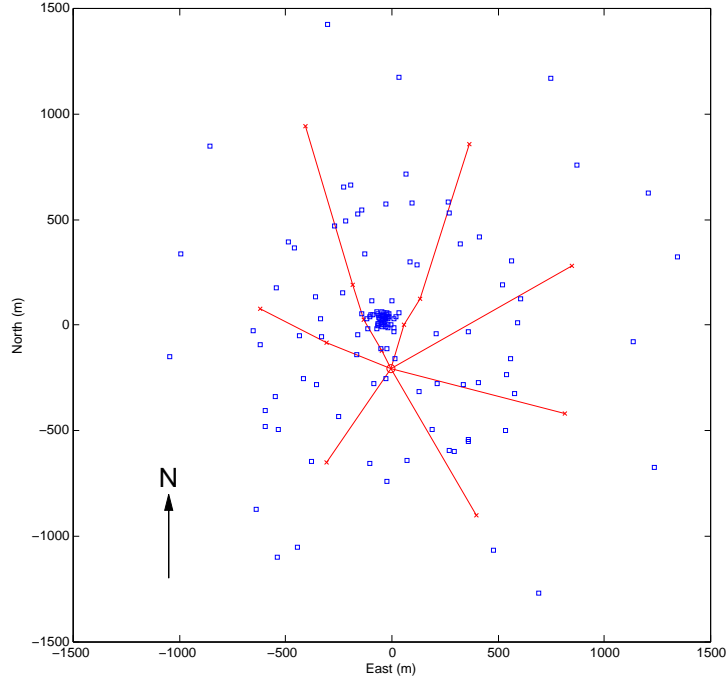}
\end{array}$
\end{center}
\caption{Plan of MWA infrastructure, as built at the MRO.  The hub referred to in the text is located at (0,-200), with seven radial trench lines shown in red.  Receiver locations are marked as red crossess and tile positions are marked as blue squares.}\label{fig:infrastructure}
\end{figure*}

The hub, located approximately 200 m south of the MWA core, is the central distribution point for the power and data services into, from and around the array (Figure \ref{fig:infrastructure}); it consists of power and data distribution apparatus mounted on separate but adjacent concrete plinths.

A compact substation consisting of a transformer and a Low Voltage (LV) switchboard housed in an environmental enclosure (kiosk) is mounted on one concrete plinth. All cable entries are from underneath through apertures in the precast concrete plinth. The transformer is a three phase ONAN type (O=mineral oil or synthetic insulation fluid with a fire point 300$^{\circ}$; N=natural convection flow through cooling equipment and in windings; A=air is the external cooling medium; N=natural convection of the external cooling medium) with a ratio of 6.6kV/415V, a Dyn11 winding arrangement and an impedance of 4\%. The incoming cable enters the termination enclosure via an all-metal cable gland which is bonded to the cable sheath. The HV termination enclosure is sealed with conductive RFI gaskets. The LV switchboard comprises the main incoming terminal, including a three pole manual changeover switch (mains/off/generator) and the outgoing mains feeders. The incoming and outgoing feeders are protected by fused switches. No electronic equipment is employed within the compact substation.

A high density fibre optic patch panel housed in an environmental enclosure is mounted on a separate but adjacent plinth. All fibre entries are from underneath through apertures in the precast concrete plinth.

The MWA distribution hub is connected to the wider MRO power and data networks by a single incoming services trench. The incoming high voltage cable is a 35 mm three core armoured cable with aluminium conductor and an operating voltage range of 6.35/11kV, designed and constructed to comply with AS/NZS1429.1-2006. The incoming data cables are Prysmian Fusion link dry ribbon cables consisting of 18 ribbons of 12 fibres (216 single mode fibres) operating at 1310 nm. There are two incoming cables providing a total of 432 incoming fibres. An extruded nylon sheath provides environmental protection.

Distribution from the MWA hub is by cables direct buried in service trenches that terminate at service pillars designed to provide direct plug-in points for MWA receivers (and other hardware that can make use of the same service formats). The LV cabling has a copper conductor of various diameters based on the load and length of individual cable runs packaged in flexible cross-linked polyethylene insulation and with an extruded nylon protective barrier. The optical fibre cables consist of 12 cores of standard single mode fibre in a cable designed for direct burial and with an extruded nylon barrier. The capacity of the fibre and power deployed is designed to accommodate possible expansion of the core of the array at some future point in time.

The service pillars are equipped for single phase distribution and include a single phase distribution board used to supply either one or two receivers. The distribution board also provides a local isolation point for the receivers. The service pillars are mounted on concrete plinths with bottom cable entry facilitated by apertures in the precast concrete. The final connection to receiver is by means of a standard three-pin plug and socket.

The CPF houses the MWA post-receiver signal processing equipment within 16 Schroff LHX20 water cooled 19" cabinets (42U height $\times$800 mm width$ \times$1200 mm depth).

\subsection{Array configuration}

The initially envisaged MWA design consisted of 512 tiles distributed with a $r^{-2}$ density profile for radii of 50$-$750~m, with a flat distribution of tiles inside of 50~m and 16 outlying tiles at $\sim$1500~m radius \citep{Lonsdale-etal.2009}. This profile provided smooth $uv$ coverage with a strong concentration of short baselines at the scales relevant for EoR power spectrum measurements \citep{bea12}, the excellent snapshot imaging needed for transients and holographic calibration \citep{haz12}.

\begin{figure*}[ht]
\begin{center}$
\begin{array}{cc}
\includegraphics[scale=0.1, angle=0]{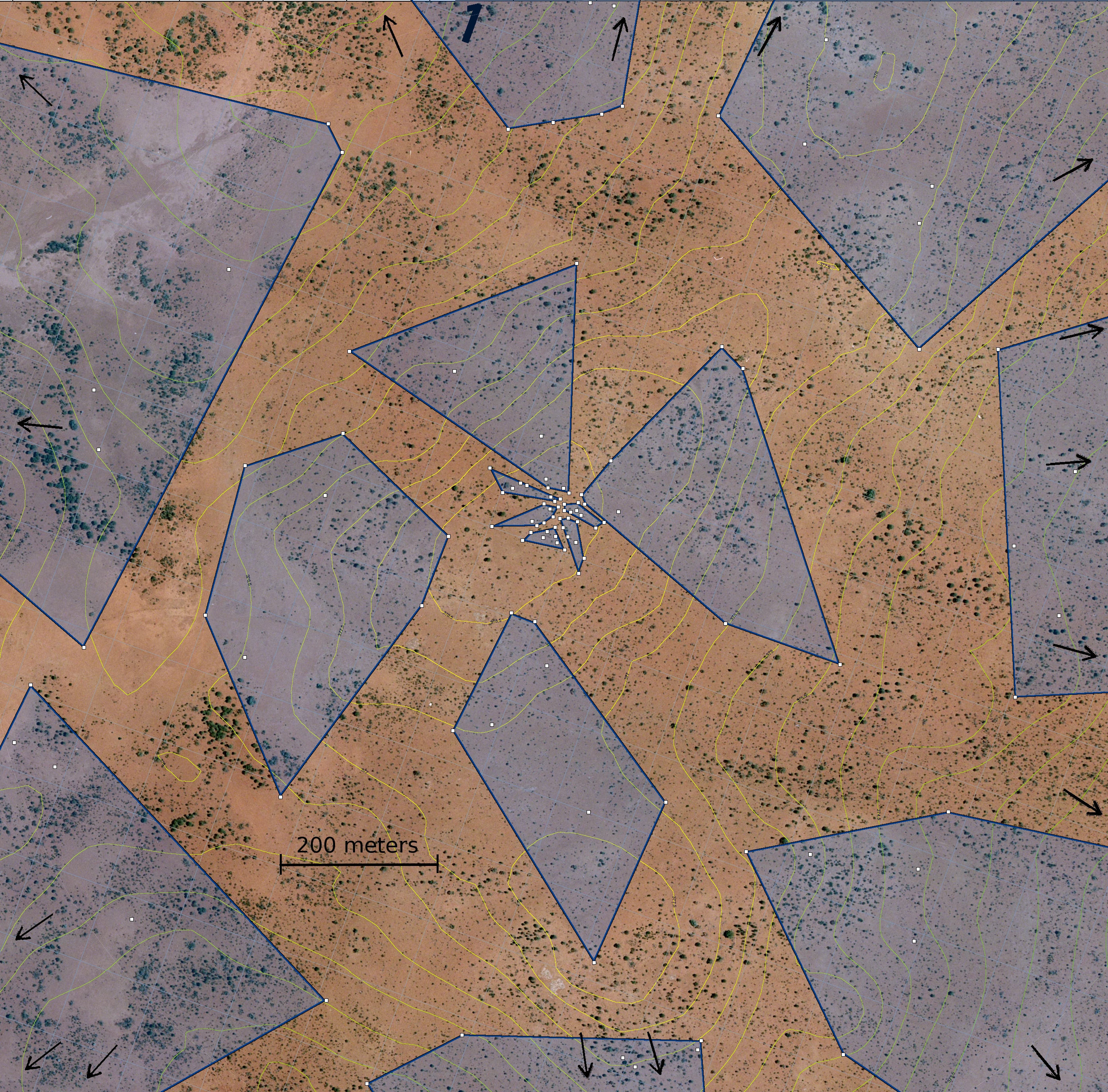} &
\includegraphics[scale=0.2, angle=0]{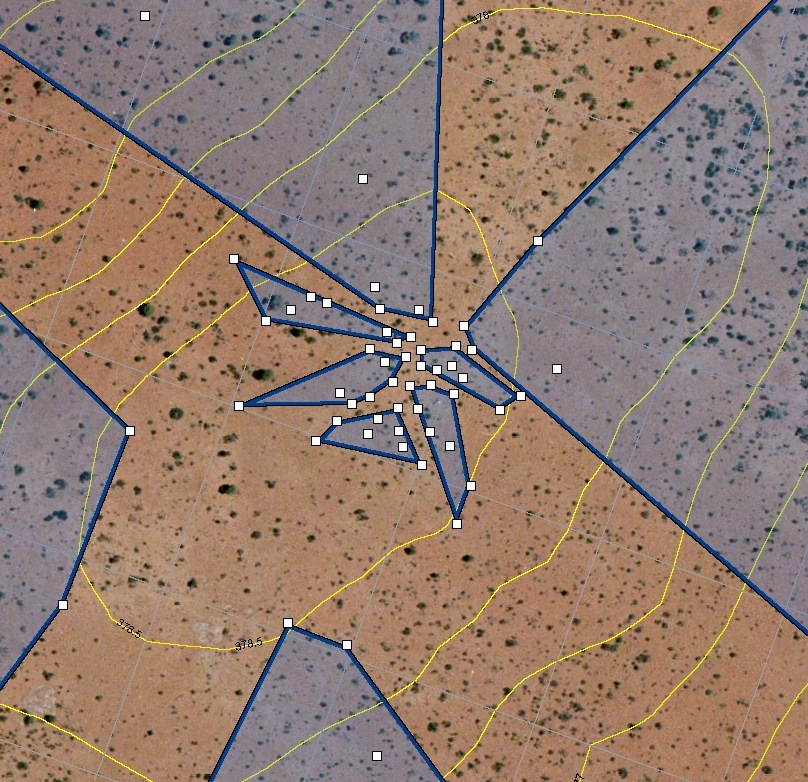}
\end{array}$
\end{center}
\caption{Left: An aerial map of the central 112 antennas of the MWA layout, indicated by white 5x5~m squares. The layout features a dense core of antennas within 50~m of the array center, with a very smooth distribution of antennas out to 750~m radius. The arrows indicate the direction to the outer 16 antennas for high angular resolution imaging which are in a rough ring of $\sim$1.5 km radius. The purple regions indicate the electrical footprints of each of the 16 receivers, each servicing 8 antennas.  Right: A zoom into the central $\sim$1 km of the array configuration.  The contours }\label{fig:layout}
\end{figure*}

\begin{figure*}[ht]
\begin{center}$
\begin{array}{cc}
\hspace{-5mm}
\includegraphics[width=0.5\textwidth, angle=0]{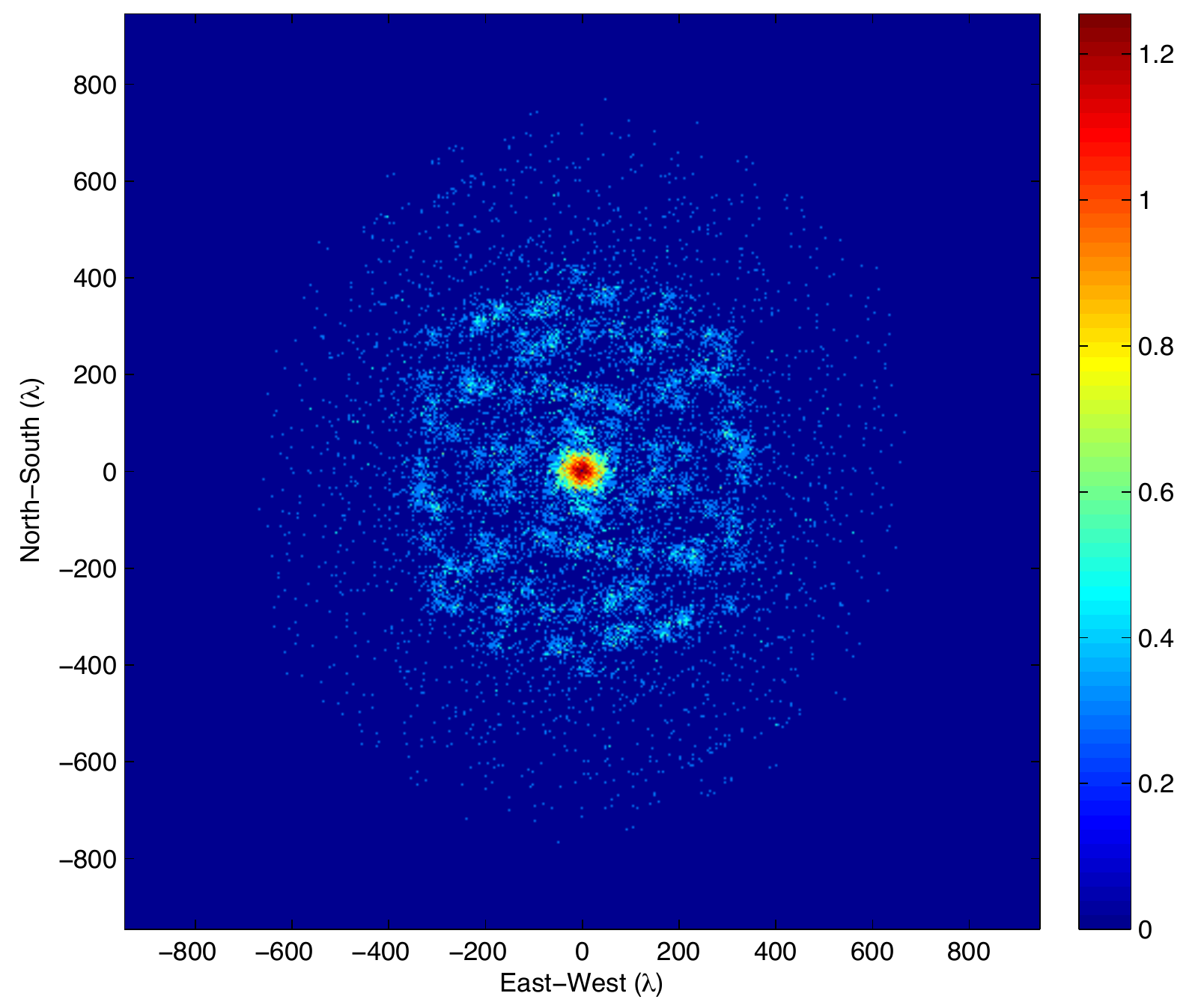} &
\includegraphics[width=0.5\textwidth, angle=0]{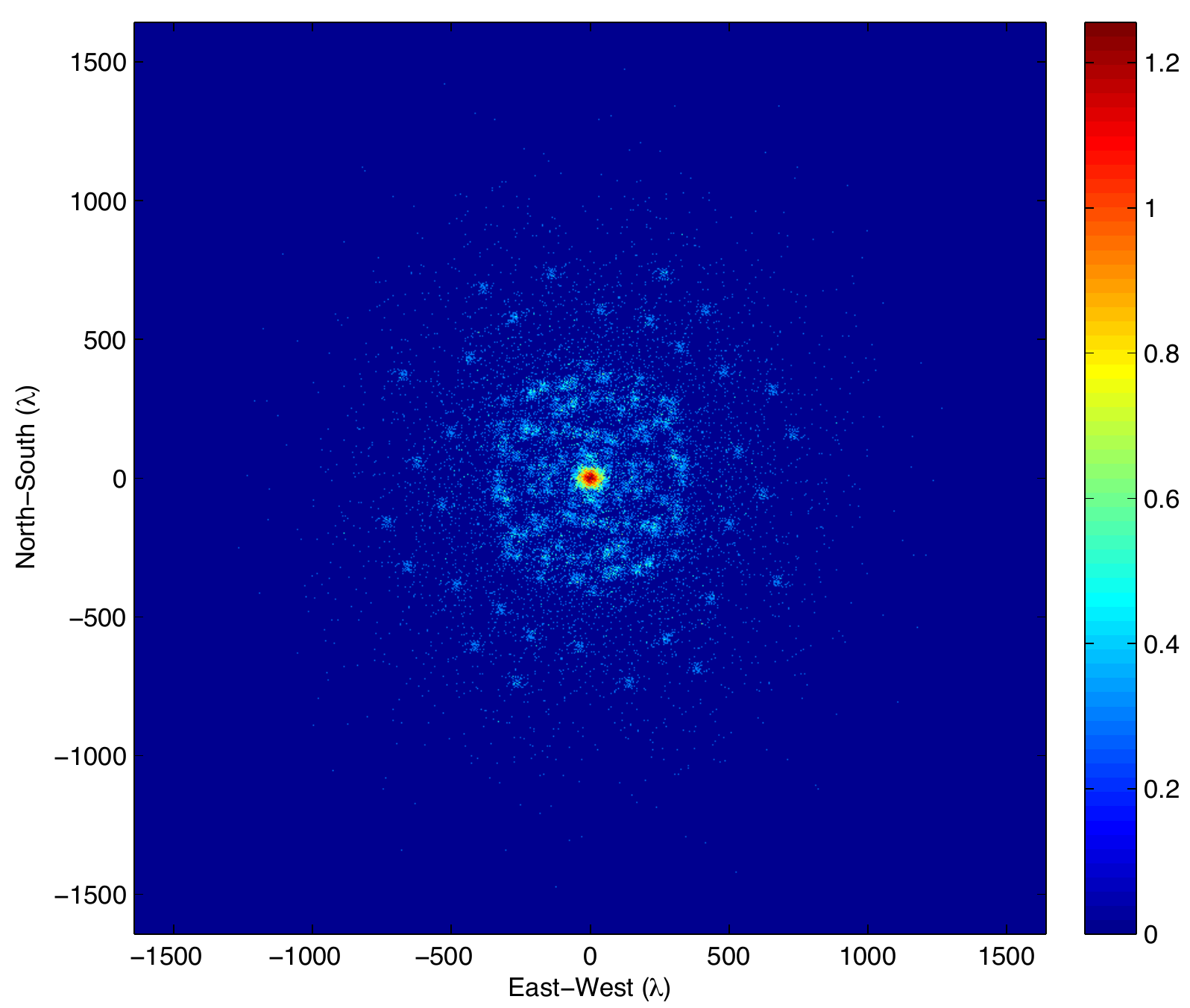}
\end{array}$
\end{center}
\caption{Left: The snapshot single frequency $uv$ coverage for the central 112 antennas.  Right: The snapshot single frequency $uv$ coverage for the full array.}\label{fig:uv1}
\end{figure*}

In rescoping the array to 128 tiles, we maximised science capacity by keeping the key figures of the original layout: excellent snapshot $uv$ coverage, a concentrated core of tiles for Epoch of Reionisation (EoR) power spectrum measurements, and longer baselines for solar observations. This has lead us to place 25\% of the tiles within a dense 100~m diameter core, with a very smooth tile distribution out to 1.5~km diameter and 16 tiles in an outer region of 3~km diameter.  The resulting tile distribution is shown in Figure \ref{fig:layout} and the snapshot single frequency $uv$ coverage is shown in Figure \ref{fig:uv1}.  The coordinates of the centre of the array are: Latitude $-$26$^{\circ}$ 42' 11.94986''; Longitude 116$^{\circ}$ 40' 14.93485''; Elevation 377.827 m.

Even in the remote Australian desert, there are areas of rock outcroppings, emergency runways, flood zones, and heritage areas where tiles cannot be placed. \citet{bea12} details a new tile placement algorithm that produces extremely smooth snapshot $uv$ distributions in the presence of ground exclusion zones. Noting that over- and under-densities in the $uv$ plane act like two dimensional wave packets to produce PSF (array beam) sidelobes, the new method uses a Bessel decomposition of the $uv$ plane to create very smooth baseline distributions. 

\subsection{Tiles and analog beamformers}

\begin{figure*}[ht]
\begin{center}
\includegraphics[scale=0.5, angle=0]{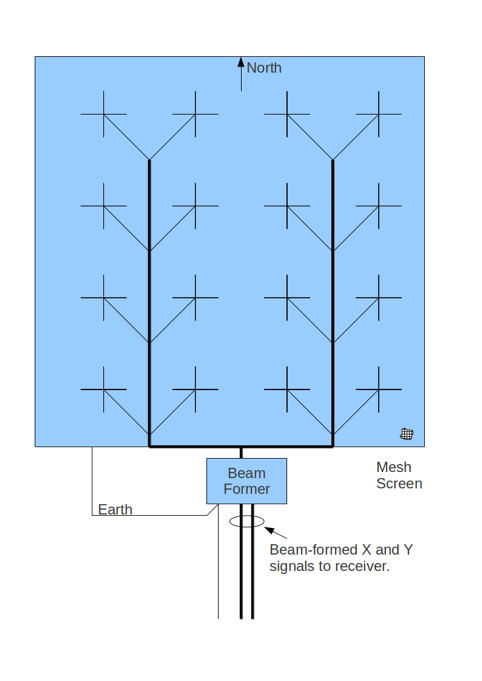}
\caption{A schematic antenna tile layout. Note that element cables are not shown to length scale.}\label{fig:antenna_schematic}
\end{center}
\end{figure*}

\begin{figure*}[ht]
\begin{center}
\includegraphics[scale=1, angle=0]{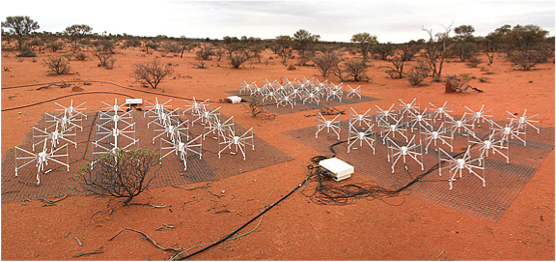}
\caption{MWA tiles and analog beamformers deployed in the field.}\label{fig:antenna_real}
\end{center}
\end{figure*}

The MWA antenna system is composed of the active and passive components described below. These components, when connected together, form a ``tile" and analog beamformer (Figures \ref{fig:antenna_schematic} and \ref{fig:antenna_real}). The MWA has deployed 128 antenna tiles and beamformers, in a configuration described in Section 2.2.

A 5 m $\times$ 5 m mesh reflecting screen (effectively a plane mirror over the MWA operating frequency range), consisting of galvanised steel wire mesh with 50 mm $\times$ 50 mm wire spacing, and 3.15 mm wire thickness, forms the ground screen for each tile. These ground screens rest directly on the ground, each positioned to align within 2$^{\circ}$ of north-south/east-west, within 1.5$^{\circ}$ of perpendicular to the zenith, and with less than 5 cm deviations from planarity across each ground screen.  The ground screen is composed of three rectangular sheets spot welded together sufficiently well to provide a continuous electrical path and is electrically connected to the metal chassis of the analog beamformer which, through a dedicated drain wire, is connected to the metal chassis of the receiver, ensuring a good discharge path for static electricity.  

Sixteen dual polarisation active antenna elements form a 4 $\times$ 4 regular grid on the ground screen.  Each antenna element consists of two sets of orthogonally mounted aluminium broad-band dipole ``bat-wings", each pair directly feeding a custom designed Low-Noise-Amplifier (LNA) circuit. The LNA circuit provides approximately 19 dB of gain at 150 MHz when terminated at 50 ohms, and includes a feed conversion from balanced to single-ended. The LNA circuit is powered from a 5~VDC bias supplied by the analog beamformer. Antenna elements are mounted on the mesh with one dipole aligned north-south and the other east-west. While these active dipoles are broad-band by design, the anti-aliasing filter immediately prior to the digitisation stage in the receiver limits the telescope operating range to approximately 80$-$300 MHz. The centre-to-centre element spacing on the mesh is 1100 mm corresponding to half wavelength separation at 136 MHz.

\begin{figure*}[ht]
\begin{center}
\includegraphics[scale=0.25, angle=0]{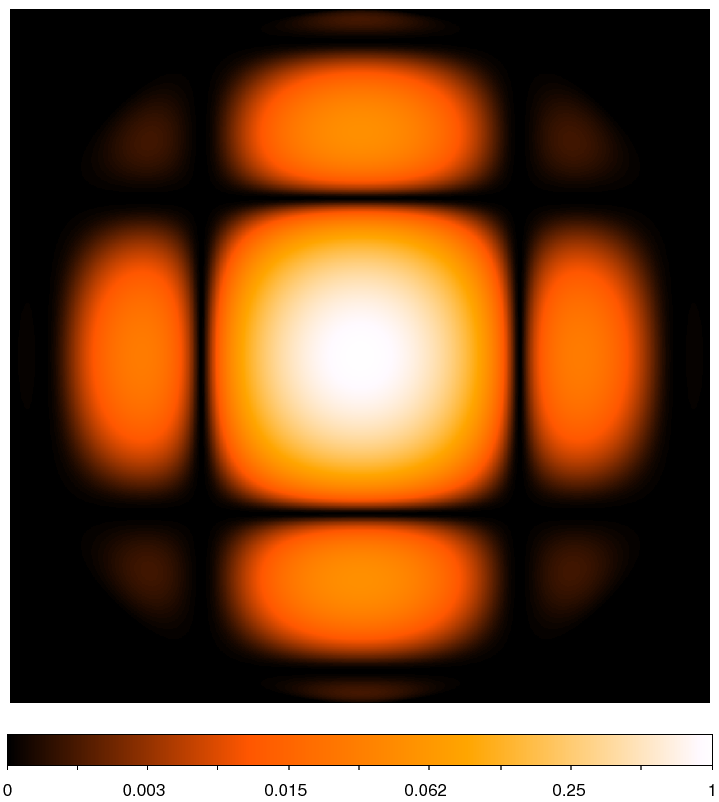}
\caption{Simulated beam response for a single MWA tile, as described in the text.}\label{fig:beamshape}
\end{center}
\end{figure*}

The signals from the 16 $\times$ 2 dipole antennas are wired to the analog beamformer using 50-ohm cable specified to LMR-100. All these element cables are made to the same length of seven metres to ensure that all antenna signals reach the analog beamformer front panel with identical delays. As mentioned above, these cables also carry the 5VDC bias from the beam-former to power the LNA circuits.  Each of the 32 channels in the analog beamformer has an independently controlled 32-step delay using five binary-weighted delay steps, each of which can be switched in or out. When any delay step is switched out, a gain-matching circuit stub is switched in to maintain relatively constant gain regardless of delay setting (tile gains are calibrated post-correlation). The delay steps are multiples of 435 picoseconds, allowing a range of discrete relative delays between elements, from zero to 13.5 ns. The outputs from all 16 delay stages for each polarisation are summed to create X- and Y-polarised tile beams on the sky. The analog beamformer, and hence the instrument, can support steering the X- and Y-beams for all 128 tiles in different directions, for a total of 256 beam directions. It should be noted that the beam shape changes with pointing direction (as well as frequency) and pointing directions below 30$^{\circ}$ elevation are not available. Figure \ref{fig:beamshape} shows the simulated beam response of a single MWA tile, at a frequency of 150~MHz and pointed at zenith.


The summed X- and Y-outputs are amplified and impedance converted to match the 75-ohm dual coaxial cable that connects the analog beamformer to the receiver. This cable also carries encoded control signals to set the delays on the analog beamformer, as well as the 48 VDC power for the beamformer. Shorter cables conform to RG-6 specification, while longer cables meet LMR-400-75 specification. Furthermore, on longer cable runs a ``whitening" filter is inserted to overcome some of the frequency-dependent loss that would otherwise over-emphasise the low-frequency signals.

\subsection{Receivers, digitisation and first stage channelisation}

\begin{figure*}[ht]
\begin{center}
\includegraphics[scale=0.5, angle=0]{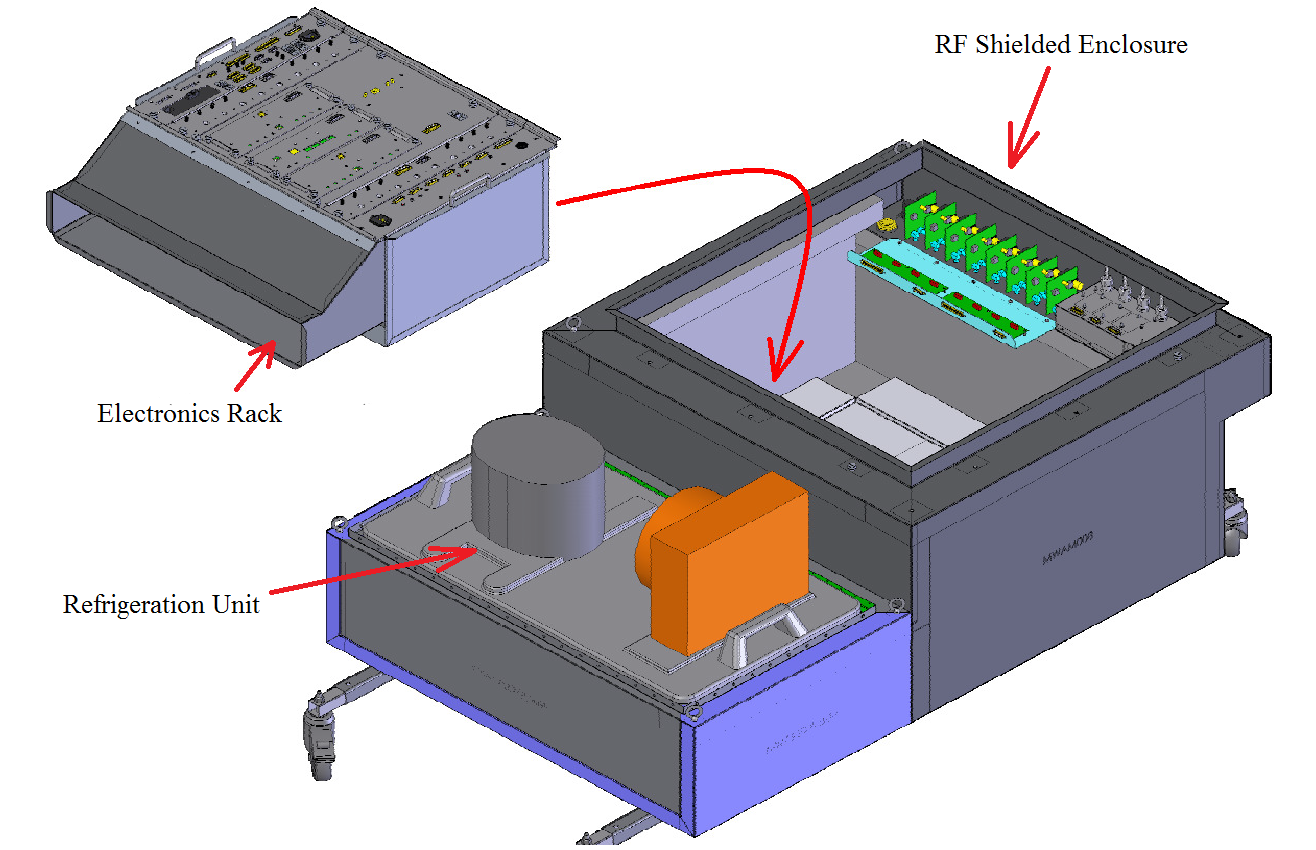}
\caption{Exploded view of an MWA receiver package.}\label{fig:rx_ex}
\end{center}
\end{figure*}

An MWA receiver node unit (Figure \ref{fig:rx_ex}) is responsible for taking the analog radio frequency (RF) signals from 8 tiles, performing digitisation and coarse frequency channel selection of these signals, and then transmitting the resulting digital streams via fibre optic cable to the CPF.

There are a total of 16 receiver nodes in the complete 128 tile MWA system.  These receivers are physically distributed around the array to minimise the length of the coaxial cables that carry the analog RF from the beamformers to the receivers.
synchronisation
RF signals arriving at the receiver are passed to an Analog Signal Conditioning (ASC) board where signal-level adjustment, impedance matching and 80$-$300 MHz band-pass filtering takes place.  The conditioned analog signals are then passed to an Analog-to-Digital and Filter Bank (ADFB) board.  The signals are sampled at 8 bits per sample and 655.36 Msamples/sec by ATMEL AT84AD001B ADC chips and fed to Xilinx Virtex 4 Field Programmable Gate Array (FPGA) chips.

The FPGAs implement a 256 channel coarse polyphase filter bank which gives 256$\times$1.28 MHz wide channels over the 327.68 MHz sampled bandwidth.  Of these 256 coarse channels, channel numbers below about 55 and above about 235 are highly attenuated by the ASC filtering.

The channelised data are transferred by a custom designed high-speed backplane to the Virtex 5 FPGA based Aggregator Formatter (AGFO) board.  A user-defined subset of 24 of these channels (not necessarily contiguous) are formatted and transmitted on three fibre optic cables.  Each of the three fibres contains data for 8 of the selected coarse channels for each of 2 polarisations for the 8 tiles connected to that receiver.  The three fibres together yield 30.72 MHz of RF bandwidth that is transmitted to the CPF.  The data are transmitted in the form of 5+5 bit complex samples.  Together, the 16 receiver nodes in the 128 tile system pass in excess of 80 Gbits/s of data to the down-stream signal processing systems.

Additional fibres provide Ethernet communications for monitor and control functions and distribute a centralised clock signal for the samplers, FPGA logic and timing signals for array synchronisation. A single-board computer controls the receiver node functions and services monitor and control (M\&C) needs. Control functions include analogue beamformer commands, setting and monitoring of ASC units, configuring and monitoring of high-speed digital boards, thermal control and managing power start-up and shut-off in response to various conditions.

The receiver electronics are fitted into a rack which is housed in a weather-tight, RF shielded enclosure with an integrated refrigeration unit.  This allows the receiver to meet the stringent radio quiet requirements of the MRO and to provide environmental conditioning and protection to enable the unit to survive in an environment where the ambient air temperature can range from 0$^{\circ}$ to 50$^{\circ}$C.

\subsection{Correlation}

The real-time cross-correlation task for interferometric arrays is a large computational challenge, historically addressed by  Application Specfic Integrated Circuits (ASICs) and FPGAs. These technologies are well matched to the limited precision fixed point arithmetic required, but their application typically involves trade-offs between cost of development, power consumption and performance. ASICs are costly to develop and produce; FPGAs are cheaper to develop than ASICs, and both are much more power efficient than general purpose processors. However, as discussed by \citet{NieuwpoortRomein}, there has been considerable recent effort expended applying many-core processors to this problem. The MWA has chosen to leverage two technologies to perform the correlation task, a polyphase filtering task (PFB)  performed by an FPGA-based solution, and a cross-multiply and accumulate task (XMAC) utilising GPGPU technologies developed by \citet{Clarketal}.

The purpose of a spectral line correlator is to measure the level of signal correlation between all antenna pairs at different frequencies across the observing band. The result is commonly called the {\em cross power spectrum} and for any two antennas $V_{1}$ and  $V_{2}$ can be formed in two ways. Firstly the cross correlation as a function of lag, $\tau$, can be formed:
\begin{equation}
(V_{1} \star V_{2})(\tau) = \int_{-\infty}^{\infty} V_{1}(t)V_{2}(t-\tau) dt.
\label{eqn:xcorr}
\end{equation}
The cross power spectrum, $S(\nu)$, is then obtained by application of the Fourier transform to reveal:
\begin{equation}
S_{12}(\nu) = \int_{-\infty}^{\infty} (V_{1} \star V_{2})(\tau)e^{-i2\pi\nu\tau} d\tau.
\label{eqn:XF}
\end{equation}
When the tasks required to form the cross power spectrum are performed in this order (lag cross-correlation, followed by Fourier transform) the combined operation is considered an XF correlator. However the cross correlation analogue of the convolution theorem allows Equation \ref{eqn:XF} to be written as the product of the Fourier transform of the voltage time series from each antenna:
\begin{equation}
S_{12}(\nu) = \int_{-\infty}^{\infty} V_{1}(t) e^{-i2\pi\nu t} dt  \times \int_{-\infty}^{\infty} V_{2}(t) e^{2\pi\nu t} dt.
\label{eqn:FX}
\end{equation}
Implemented as Equation \ref{eqn:FX}, the operation is described as an FX correlator. The FX correlator has a large computational advantage. In an XF correlator the cross correlation for all baselines requires O(N$^{2}$) operations for every lag, and there is a one to one correspondence between lags and output channels, F, resulting in O(FN$^{2}$) operations to generate the full set of lags. The Fourier transform requires a further O(F$\log_{2}$F) operations, but this can be performed after averaging the lag spectrum and is therefore inconsequential. For the FX correlator we require O(N$\log_{2}$F) operations per sample for the Fourier transform of each data stream, but only O(N$^{2}$) operations per sample for the cross multiply (although we have F channels the sample rate is now lower by the same factor). As discussed by \citet{Clarketal} there is a huge computational advantage in implementing an FX correlator of at least a factor of F. However XF correlators have been historically favoured by the astronomy community, at least in real-time applications, as there are disadvantages to the FX implementation. The predominant disadvantage is data growth: the precision of the output from the Fourier transform is generally larger than the input, resulting in a data rate increase and there is also the complexity of implementing the Fourier transform. 

\begin{figure*}[ht]
\begin{center}
\includegraphics[scale=0.5, angle=0]{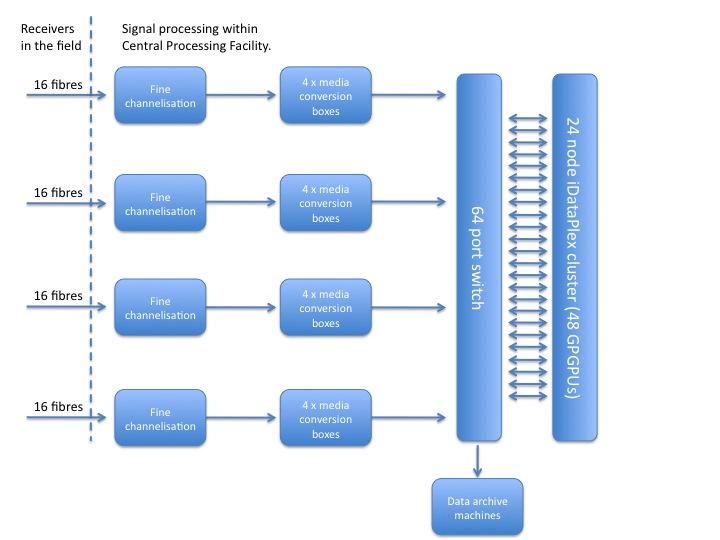}
\caption{Correlation sub-system overview}\label{fig:corr_overview}
\end{center}
\end{figure*}

The MWA implements a hybrid, distributed FX correlator solution that efficiently deals with the disadvantages of an FX correlator (Figure \ref{fig:corr_overview}). The F-stage is performed in two stages, first as a coarse, complex filter in the receiver, as described in the previous Section, and then to finer spectral resolution by 4 dedicated FPGA-based polyphase filter bank (PFB) boards co-located with the cross-multiply system in the CPF. The fine channelisation system captures 48 fibres in total from 16 receiver nodes; each fibre carries 24 x 1.28 MHz spectral channels from both polarisations of 8 antenna tiles digitised to 5~bit precision. An individual PFB board ingests the full bandwidth from 4 receivers (32 tiles) and performs a further factor of 128 in channellisation resulting in 3072 x 10 kHz channels. The output of each PFB is therefore 3072 Nyquist sampled, 10kHz channels from 2 polarisations of 32 tiles, presented on serial data lines via an interface module known as the rear transition module (RTM). The last stage of the PFB is a bit selection stage that selects only the most significant 4 bits to represent the samples. This effectively counters the major disadvantage of the FX correlator, data growth. In the MWA case, as 5 bit precision is input, and 4 bit precision is output, the F stage performs a slight reduction in data rate. 

The MWA cross-multiplication and accumulate (XMAC) is frequency multiplexed and performed independently on 24 IBM iDataPlex dual Xeon servers, each housing 2 $\times$ NVIDIA Tesla M2070 Graphics Processing Units. Each machine is allocated 128 contiguous 10 kHz channels from all the antennas. A Chelsio 10 GbE interface card and a 64 port IBM BNT 8264 switch are used to provide the packet switching required to aggregate the data from each PFB output. To permit packet switching the output of the PFB must be converted from the Xilinx propriety serial protocol (RocketIO) into 10 GbE. The media conversion is performed by a set of 16 dedicated dual Xeon servers, each housing a Xilinx FPGA-based capture board (supplied by EDT Incorporated) which accepts a portion of the serial data from the PFB and presents it to system memory where a bespoke software application routes each packet to a target machine. A connection based protocol (TCP) is used due to the complex nature of the packet switching and the synchronisation issues arising from capturing many parallel streams in a general purpose computing environment. 

Although the purpose of the operations described above is to facilitate re-packetisation from RocketIO to 10 GbE, the 16 servers also provide a capability to capture the voltage samples directly to disk. The voltage capture system (VCS) will record a copy of the voltages using a continuous ring buffer via a SAS-2 controller in each server with enough 10,000 rpm hard drives to store an hour of data at minimum.  Upon receiving a trigger, the ring buffers will form new rings using unused memory on the drives, thus preserving the data from the previous buffers.  The VCS servers contain enough memory to store at least 4 minutes of data which can be searched through for transients with inherent timescales shorter than the integration time of the telescope using incoherent dispersion and to prompt the trigger on the ring buffer when a detection occurs. Alternatively, an external trigger can be used to save the data from the ring buffer. Room has been intentionally reserved within this system to incorporate real-time GPU processing at a later date.  The recorded voltages can be correlated using existing software correlation solutions, such as DiFX \citep{del07,del11}.
	

In order to obtain data from all the tiles  for its frequency allocation, each of the 24 XMAC servers has to accept packets from all of the media converters, and obtain them in the correct order within a narrow time interval. This task is enabled by extensive use of parallel programming methodologies. Despite the large aggregate bandwidth, the bandwidth per XMAC machine is only 2.8 Gbps, and is well within the PCIe2.0 bandwidth of 64 Gbps available to the 16-lane Telsla M2070. The GPU enabled XMAC operation is described in detail by \citet{Clarketal} and the level of computation required to processes 1.28~MHz of bandwidth for 128 tiles is 335 GFLOP which is also well within the capability of the XMAC kernel we employ. The XMAC is fully performed within the GPU environment, with an output resolution of 0.5 s, and a channel resolution of 10 kHz (which will be combined to 40 kHz to facilitate storage). The format is half the correlation matrix for each channel with a lightweight FITS \citep{Wellsetal1981} header and the output data rate from the XMAC is 3.2 Gbps.

\subsection{Real-time imaging and calibration}

The ever-growing data rates generated by next generation radio arrays, in the MWA case driven by the wide fields of view and correlation rich architecture, are forcing astronomers to integrate visibilities over longer time intervals than are otherwise desirable.  One approach to dealing with these problems is to make snapshot images and to extend data reduction into the image domain, where various types of direction-dependent corrections can be readily made.  

\begin{figure*}[ht]
\begin{center}
\includegraphics[width=\textwidth, angle=0]{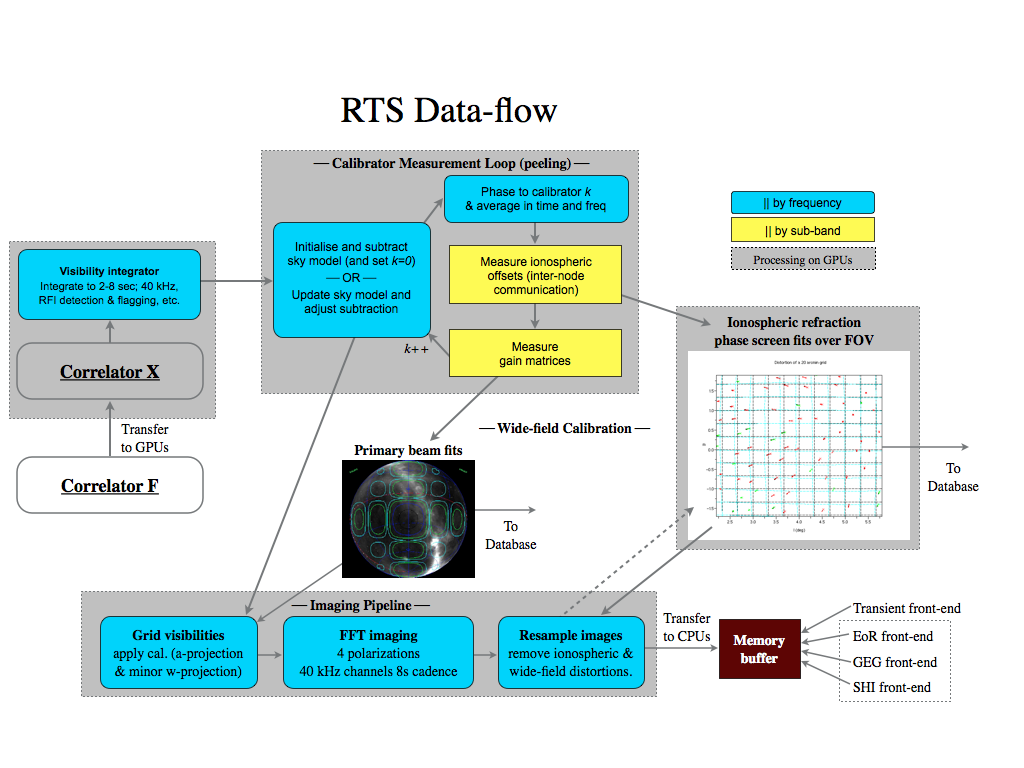}
\caption{Overview of MWA RTS, as described in text.}\label{fig:real-time}
\end{center}
\end{figure*}

The MWA sub-systems outlined in previous Sections contain novel elements, such as wide-field fixed dipole antennas and GPU-based software correlation, but describe signal chain operations that are largely traditional in nature.  Real-time imaging and calibration systems at the correlator output are emerging as a critical element of next generation wide field of view arrays, for example as demonstrated for LOFAR \citep{lofar} and as planned for ASKAP \citep{jon08,jon07}.  As such, we include below a relatively extensive discussion of the MWA approach to this sub-system (the Real-Time System: RTS) and function.  Figure \ref{fig:real-time} presents a schematic overview of the following description.

Synthesis imaging is centred around inverting the three-dimensional van Cittert-Zernike theorem, which describes the transformation between the desired sky brightness distribution and the measured visibilities \citep{Thompson-etal.2001}.  Solutions to this problem for arrays like the MWA are challenging, due mainly to four issues that worsen with increasing field-of-view (FOV) size: how best to deal with the three-dimensional problem in a way that is computationally feasible and conducive to high-dynamic-range deconvolution; how best to deal with an ionosphere that causes time- and direction-dependent distortions; how best to deal with antennas that have primary beam and polarised feed configuration variations in time, frequency and direction, and that differ from one antenna to the next; and how best to deal with the large data rates and data volumes. For an approximately coplanar array like the MWA, an imaging approach based on snapshots can meet all of these challenges: one can think of snapshot visibilities as a slice through the three-dimensional visibility set, for which the ionosphere and the nominal feed configuration matrices toward each source are constant. They are also consecutive in time and so are well suited to pipeline processing. We will consider each of the issues separately.

For a sufficiently short snapshot observation, and at the expense of a non-uniform sky brightness coordinate distortion, the transformation between sky brightness and the visibilities of a coplanar array reduces to two dimensions \citep{Bracewell1984,CornwellPerley1992,Perley1999,Cornwell2005,Ord-etal.2010}. The non-uniform coordinate distortion comes from imaging on a plane that is parallel to the array plane, rather than the tangent plane at the field centre, and thus it will change as the field centre is tracked. Once these coordinate distortions have been removed, which requires an image re-sampling process, time integration can continue in the image domain after appropriate pixel-wise weighting to maximise the signal-to-noise ratio. This is known as \emph{warped-snapshot imaging}, and in the RTS the procedure is combined with ionospheric corrections and conversion to right ascension and declination coordinates \citep[in the HEALPIX pixelisation scheme;][]{Gorski-etal.2005}.

Averaging in time is computationally expensive when it is done in the image domain, and in many traditional situations the pixel-based operations will dominate the processing, often by orders of magnitude. For arrays like the MWA, however, as the number of visibilities grows relative to the number of pixels, it can become competitive or even cheap compared to alternatives such as W-projection \citep{Cornwell2005}. On the other hand, removing the non-uniform warp from the snapshot images will lead to a shift-variant point spread function (PSF), which can significantly effect deconvolution, as described by \citet{Perley1999}. However, the variable primary beams and the variable atmosphere described below, which are realities for most low-frequency dipole arrays, can both lead to variable PSFs, both in position and time, and joint-deconvolution approaches will likely be required, regardless of the imaging strategy.

One of the most compelling reasons to make snapshot images is the ionosphere. During good ionospheric conditions, the size scales of the ionosphere that are expected to be relevant for the MWA are large relative to MWA baselines lengths \citep{Lonsdale-etal.2009}. The effects of the ionosphere in such a situation reduce to direction-dependent shifts in source positions, and direction-dependent Faraday rotation that is constant across the array. While it is possible to deal with the time-dependent ionospheric distortions in the Fourier domain, the angular variations expected for the MWA will require either very large gridding kernels or many image facets. In a warped-snapshot imaging approach, position distortions can be removed during the warped pixel re-sampling step, while ionospheric Faraday rotation can be taken into account during pixel-by-pixel polarisation conversion. In the RTS, the distortions are modelled by interpolating between position measurements of many compact sources with known positions, at the imaging cadence (i.e., every 8 seconds). It is unclear how well we will be able to measure ionospheric Faraday rotation, but an array of GPS receivers is being tested for this purpose, and we note that we only need to correct changes in ionospheric effects that occur within the image averaging time interval (i.e., several minutes). Any information that becomes available later (by carefully re-processing the data for polarised sources, from other instruments, etc.) can be incorporated during post-processing.

Primary beams that differ from visibility to visibility (due, for example, to changes in time or differences between antennas) are not just of concern for low-frequency dipole arrays. They are a part of wide FOV imaging where high dynamic range is required. Making direction-dependent corrections at the sample level is possible if one incorporates primary beam information into the visibility gridding kernels, an approach that is being investigated by a number of groups \citep{Myers-etal.2003,Bhatnagar-etal.2008,MoralesMatejek2009,Smirnov2011,Mitchell-etal.2012}. These approaches typically require larger gridding kernels, which can be computationally demanding. This type of gridding also results in the pixel-wise weighting required for optimal snapshot integration \citep{Ord-etal.2010}. Arrays of dipoles have an added complication, however, in that the Jones matrices that convert between sky and instrument polarisations are both time- and direction-dependent \citep{Mitchell-etal.2012}. These transformations are difficult to correct for in Fourier space when large FOVs are involved, and will also lead to large convolution kernels or will require many facets on the sky. The RTS approach is a good fit to this problem, since any direction-dependent coordinate corrections can be made in the image domain. When generating visibility gridding kernels, the RTS uses a unique primary beam model for each snapshot, frequency, polarisation and antenna. It also has the ability to update the models in real time, using gain measurements generated while peeling strong, compact sources.

Finally, a few points should be made about real-time averaging of gridded data. The RTS will average images over several minutes, storing the resultant integrated images. During each interval, the ionospheric and coordinate variations described above are typically fairly small, and the larger variations that occur over hours can be dealt with in post-processing. Furthermore, since we grid for a direction that is normal to the plane of the array, the size of the gridding kernels can be kept at a minimum. This is not the case for approaches cited above, in which most of the corrections occur during gridding, and so would need to occur in real time. The task is still extremely computationally demanding, and the RTS has been designed and written to run on high-performance GPUs \citep{Edgar-etal.2010,Ord-etal.2009}.

From the discussion above, it should be clear that the real-time calibration system has two main tasks: it needs to measure primary beam patterns, and it needs to measure ionospheric distortions across the field of view. To avoid pixelisation effects during deconvolution, it will also subtract strong compact sources from the visibilities, before gridding (peeling, as described by \citet{Noordam2004}. Each GPU will process and make images for a small number of adjacent frequency channels, as described in detail by \citet{Mitchell-etal.2008} and \citet{Edgar-etal.2010}, and the baseline system for real-time processing at MWA is 32 consecutive 40 kHz channels.

Twice per second, a complete set of visibilities arrive from the correlator, split over 24 iDataPlex compute nodes in sub-bands of 1.28 MHz (the RTS is implemented on the same machines as the correlator cross-multiplication, sharing the GPU resources between the correltor and RTS).  The visibilities for each baseline are averaged in time and frequency over the longest intervals for which decorrelation from delay and fringe rates remains below one percent, which is nominally 40 kHz and either 2, 4 or 8 seconds, depending on the baseline length. After averaging, the RTS performs a series of standard calibration tasks, such as applying cable corrections, automatic RFI detection \citep[as described by ][]{Mitchell-etal.2010} and flagging, and then the visibilities are sent to the GPUs. The data are now ready to be used to solve for the instrumental and atmospheric gains and phases, a non-linear optimisation problem that we approach via peeling. For the MWA, much of the visibility noise comes from the sky itself, and the first step is to generate model visibilities from antenna primary beam models and a sky model, which are subtracted to generate residual visibilities. At minimum the sky model will include all of the strong compact sources, which will be added back in turn\,--\,ranked based on the amount of power they contribute to the visibilities\,--\,for the following calibration steps:

\begin{enumerate}

\item \emph{Rotate visibilities}: The visibility phases are rotated to be centred at the estimated calibrator position, and the visibilities for each baseline are averaged across all available time and frequency samples (i.e., averaged to 8 seconds and 32$\times$40 kHz), into a temporary working visibility set.

\item \emph{Ionospheric measurements}: A position offset for the calibrator is modelled as a baseline- and frequency-dependent offset in the imaginary part each visibility, as in \citet{Mitchell-etal.2008}, which will be zero-mean noise if the source is at phase centre. A single pan-frequency measurement is made, to help isolate ionospheric phases from any uncalibrated instrument phases.

\item \emph{Instrumental gain measurements}: A Jones matrix for each antenna is determined toward the calibrator, based on the least squares approach discussed by \citet{Hamaker2000} and \citet{Mitchell-etal.2008}. At present a running-mean is used to reduce noise in the measurements, however an improved approach that uses a Kalman filter is under investigation. For the most dominant calibrator, the optimisation is repeated for each frequency channel, and polynomial fits to the resulting Jones matrices are used for bandpass calibration.

\item \emph{Source subtraction}: If the gain and ionosphere measurements pass a set of goodness-of-fit tests, they are used to peel the source from the full resolution visibility set, thus updating the initial source subtraction. Otherwise the initial source subtraction is repeated.
        
\item If there are more calibrators in the list, loop back to step 1.

\end{enumerate}

The output of this loop, apart from the peeled visibilities, is a set of ionospheric offset measurements and a set of Jones matrix measurements for each tile, distributed in angle across the field of view and side-lobes. The ionospheric measurements are used to adjust the pixel boundaries used in the warped pixel re-sampling step, and we currently use a moving least squares approach to interpolate between the calibrators. For isoplanatic patch sizes of $\sim4^\circ$ and image sizes of $\sim30^\circ$, we will need at least 50 or 60 sources in the field to describe the phase variation of each patch. Optimally, we would like to oversample these variations by a factor of at least a few, and if needed we will make position measurements in the images themselves to increase the number of sources. A Levenberg-Marquardt approach is used to fit primary beam models, the form of which will be finalised during commissioning. We expect to make Jones matrix calibration measurements for at least a few dozen sources at the imaging cadence.

For the MWA and other similar arrays, the process known as \emph{deconvolution} is not really deconvolution, since the PSF changes with time, position, and polarisation. However, with various modifications, many of the standard deconvolution approaches can be used to accurately remove PSF side-lobes, and developments are being made on a number of complementary
fronts, see for example \citet{Pindor-etal.2011}, \citet{Bernardi-etal.2011}, \citet{Williams-etal.2012}, \citet{Sullivan-etal.2012}, \citet{Mitchell-etal.2012} and \citet{Bernardi-etal.2012}.

Finally, one of the biggest risks involved with real-time calibration and imaging is that the visibility data volume is too large to store. The ability to loop back to the raw visibilities during deconvolution is at the heart of traditional high-dynamic-range synthesis imaging and is lost in such a system. The MWA approaches high-dynamic-range imaging from a non-traditional angle, that of a densely filled aperture that has a high quality instantaneous PSF with low side-lobes. This is important for both deconvolution and real-time calibration, and also allows us to limit the amount of PSF variation from snapshot to snapshot. 

However, it is unclear how well this approach will compare with more traditional self-calibration approaches, and to ensure the best quality images, the MWA project will also store full sets of raw visibilities. Comparisons of image quality will be made using the RTS running in real time, using the RTS running on stored data in a full self-calibration and deconvolution loop, and using other imaging strategies available in standard software packages.  The RTS development and evaluation will make the MWA an important step in the development of low-frequency radio astronomy in the lead-up to the SKA.

\subsection{Offsite data transport and the MWA data archive}

The output of the RTS (calibrated images) and the visibility output of the correlator, plus miscellaneous control data and calibration information, totals approximately 4.4 Gbps at maximum data rate.  These data are streamed to a dedicated 10 Gbps connection implemented on the CSIRO optical fibre network between the MRO and Geraldton, which is seamlessly integrated onto an AARNet network dedicated to carrying data traffic from the MRO to Perth.  The termination point of the connection in Perth is iVEC, which hosts the Pawsey HPC centre for SKA Science (aka the Pawsey Centre), a new \$A80M supercomputing centre.  Up to 15 PB of data storage has been allocated at the Pawsey Centre, ramping up over a five year period from 2013, to host the MWA data archive (both images and visibility data).  At the full data rate, this storage allowance corresponds to approximately 300 $\times$ 24 hrs of observation.  In practice, some percentage of the MWA observations will be at a significantly lower data rate, thus corresponding to more than 300 days of observations.  Since the visibility data will be archived, it will be possible to reprocess visibilities in a post-observation mode, using Pawsey Centre or other resources (organised by users) and the RTS or other software packages.  This will allow gradual refinement of imaging and calibration techniques and the highest quality data products.

The Pawsey Centre is currently under construction.  However, the MWA data archive has been specified and a prototype archive system has been implemented for testing and to support commissioning and early operations.  The MWA data archive will utilise the NGAS software \citep{wic07} and is implemented on a single server machine.  The server includes a storage array consisting of 24 $\times$ 2 TB drives arranged in a RAID5 configuration for redundancy. The total amount of storage is $\sim$48 TB when parity and striping are taken into account. The storage capacity is expandable up to $\sim$170 TB if required. There are 24 separate 1 Gbps links from each of the correlator/RTS nodes that are multiplexed into the storage array at a peak sustained data rate of $\sim$3.2 Gbps for visibility data and $\sim$1.2 Gbps for RTS data, combining to data rate of $\sim$4.4 Gbps while observing. This equates to $\sim$24 hours of continuous observing time at a 32 bit real and imaginary precision, 0.5 second integrations and 40 kHz course channel configuration.  Data throughput tests have shown that NGAS is capable of streaming to the RAID5 disk array at a rate of $\sim$6.4 Gbps. The remaining bandwidth will be dedicated to management links required for operations.  

The NGAS archive has been engineered to allow file formats to change over time or new data pipelines to be installed without the need to change code or modify the underlying archive software configuration in any way. A data capture application programmer interface has been developed to be the entry point into the archive that can be embedded into any application to accommodate future data pipelines.  Engineers and scientists will have access to a web based interface that is the portal into the archive. For a given set of observation search results, users will be presented with links to data files and their associated meta-data. The archive will offer the functionality to convert raw visibility data into a standard UV-FITS format. Such tools will include the capability to greatly reduce the bandwidth and processing overheads for each user. Users will have the option to access and download the raw visibility data if required. 

\subsection{Instrument monitor and control}

The large number of tiles and the ubiquity of embedded processing have led us implement a highly distributed Monitor and Control (M\&C) system for the MWA. Most interferometer M\&C systems are monolithic in design with a tight command and control paradigm for each antenna and hardware system. In contrast, the MWA M\&C system is a much more distributed design that leverages the embedded processing in each hardware system. In essence, each local hardware controller is responsible for the health and safety of that instrument system and all of the detailed hardware state changes (e.g.\ power up sequence, order of digital mixer changes to change the receiver frequency, etc.). The central portions of the M\&C are then primarily concerned with orchestrating all of the hardware for observing and commissioning and providing a central repository for all of the configuration and monitoring information (meta-data).

Part of this change is due to the ubiquity of embedded processing $-$ even power outlets now have embedded webservers $-$ so it is reasonable to expect every system that interfaces with M\&C to have the local intelligence to perform low-level operational tasks. The other driving factor is the sheer number of tiles and field hardware systems. As the number of antennas reaches into the hundreds it becomes more difficult to maintain centralised control. Instead of trying to have a single entity simultaneously control hundreds of sub-systems, we have opted for a model where many individual entities are coordinated centrally.  The central coordinator directs the distributed entities at a high level and these entities are responsible for controlling and monitoring the given sub-systems.

At the heart of the MWA M\&C system is a relational database that serves as a central repository of all system information. This database includes current and historical configuration; the planned observing schedule (by GPS second), the detailed state changes of each hardware and software system; monitoring data; and links to associate data files with the schedule, configuration, and all associated commands and state changes.

This central database allows us to accurately reconstruct the state of the array for any observation, including a large volume of debug information. For example, we record the full broadband 0$-$300 MHz spectrum for each tile every $\sim$8 seconds. Comparing the spectra of different tiles and cross-referencing with changes in pointing direction can quickly diagnose many errors, even when they only happen sporadically. The database is implemented in Postgres, and all of the other M\&C systems interact directly or indirectly with the database.

The other major M\&C systems are:
\begin{itemize}
 \item The Scheduler loads the desired observing schedule into the database. As noted below, there are many settings which can be either set specifically or delayed until run time (e.g. use all of the antennas in the list of good antennas). The Scheduler is implemented in a scriptable Python library;
 
 \item The array configuration web pages allow the entry and modification of the array configuration information. This includes the location, serial numbers, and interconnections of all the major hardware components and includes specific pages for common configuration changes (e.g.\ moving a beamformer from one tile to another);
 
 \item Each hardware and software component has a local control program. It receives commands, implements those commands at the indicated clock tick (synchronised by the array clock and PPS), and records all of its state changes in the central database;
 
 \item The Observation Controller serves as the conductor, and reads the schedule and the configuration tables of the database and sends the appropriate commands to the field hardware. All commands are sent in advance with the GPS time of when the command should be implemented to allow clock accurate commands despite the $\sim$10~ms jitter inherent in any messaging system. The amount of advance notice is part of the interface definition for each hardware to Observation Controller relationship. The Observation controller also resolves any observation-time decisions. For example, this allows the schedule to be written weeks in advance to use only the tiles the instrument team judge as good, and which tiles are actually in that list are resolved at observation time. Even within an observation, as tiles are added and removed from the list the tiles included in the observation are dynamically adjusted. The Observation Controller is implemented in Java;
 
 \item The Facility Controller serves a similar purpose to the Observation Controller, but concentrates on startup and shutdown of the array, software and firmware updates, and restarting faulty equipment. While this could have been implemented as part of the Observation Controller, the staged startup of the array (e.g.\ making sure the necessary network switches have started prior to booting a receiver, and staging the receiver startup to limit inrush current) and software updates are conceptually quite different than observation commands and we have separated the implementation of these functions. The Facility Controller is implemented in Python;
 
 \item The Monitoring Tools present the current and historical state of the array to users. All of the system information is stored in the central database, but these tools present the database information to the user in a useful way.  A few examples include: a page that lists all of the data files produced during an observation and a summary of all the receivers and tiles that were online; a zoomable plot of the temperature of each tile vs.\ time ($\sim$ambient temperature) or each receiver vs.\ time ($\sim$air conditioner performance); a plot of the broadband 80$-$300~MHz spectra during the course of an observation; results of a commissioning observation that turns on one dipole at a time for each tile to measure the complex gain of each dipole as a function of frequency. The Monitoring Tools are implemented as a diverse and growing set of web tools;
 
 \item The Police are a set of tests that determine when a system is performing within specifications. They range from very simple (is a receiver responding to commands) to quite complex (identifying unstable LNAs from the broadband spectra). The results of the Police are logged back to the database and are the basis of alerts presented by the Monitoring Tools (turning an receiver status from green to red) and can be used to dynamically adjust the array configuration (remove a tile from the good list).
\end{itemize}

Together the components listed above act in concert to enable a very flexible and robust M\&C system. To date the system has performed well, with the inherent flexibility and the volume of monitoring data serving as a great resource in commissioning and debugging the MWA. Partly because of the modular nature of the MWA, the M\&C system is inherently scalable to many hundreds of tiles and may provide a useful starting point for future arrays with large numbers of antennas.

\section{MWA performance metrics}

For the MWA as described in this paper, the time needed to reach a point source sensitivity of $\sigma_{\rm s}$ is:

\begin{equation}
t = \left ( \frac{2 k_{\rm B} T}{A_{\rm eff} N \epsilon_{\rm c}} \right )^2 \frac{1}{\sigma^2_{\rm s} B n_{\rm p}}
\end{equation}

where $k_{\rm B}$ is the Boltzmann constant, $T=T_{\rm sky} + T_{\rm rcv}$ is the system temperature, $A_{\rm eff}$ is the effective area of each antenna tile, $N$ is the number of antenna tiles, $\epsilon_{\rm c}$ is the correlator efficiency (assumed unity for the MWA), $B$ is the instantaneous bandwidth, and $n_{\rm p}$ is the number of polarisations.   For the parameters given in Table \ref{tab:system} at 150~MHz, this equation reduces to $t\approx{\rm 8\times10^4} / B\sigma_{\rm s}^2$ seconds, for $\sigma_{s}$ in mJy.

The intrinsic source confusion limit is estimated to be $\sim$10 - 20~mJy at 150~MHz at the MWA angular resolution of $\sim1 - 5'$.  Table \ref{tab:t_sensitivity} lists approximate derived sensitivities.

\begin{table*}[ht]
\begin{center}
\caption{Sensitivity of the MWA at 150 MHz}\label{tab:t_sensitivity}
\begin{tabular}{llcl}
\hline 
Property & $B$, $\sigma_{\rm s}$, ($\Theta_{\rm B}$) & Value & \\
\hline
Surface brightness sensitivity 	&	1 MHz, 1 K, (1$^\circ$) 	&	60	&	seconds \\
Point source sensitivity 		&	1 MHz, 10 mJy			&	800	&	seconds \\
Point source sensitivity		&   31 MHz, 10 mJy		&    26 seconds \\
Broadband survey speed 		&	31 MHz, 10 mJy			&	1.5$\times10^5$ &	deg$^2$/hr \\
Narrowband survey speed		&	0.04 MHz, 10 mJy		&	190	&	deg$^2$/hr\\
\hline
\end{tabular}
\medskip\\
\end{center}
\end{table*}

\section{MWA science goals}

The MWA will be capable of a wide range of science investigations. These planned investigations are described in detail by \citet{bow12}.  Here, we review the four key science themes that encompass the planned investigations and that have driven the design of the array.  The key science themes for the array are: 1) detection of fluctuations in the brightness temperature of the diffuse redshifted 21 cm line of neutral hydrogen from the epoch of reionisation (EoR); 2) studies of Galactic and extragalactic (GEG) processes based on a deep, confusion-limited survey of the full sky visible to the array; 3) time domain astrophysics through exploration of the variable radio sky (transients); and 4) solar heliosphere and ionosphere (SHI) imaging and characterisation via propagation effects on background radio sources.

Exploration of the Cosmic Dawn, the period when the first stars and galaxies formed in the early Universe, has been identified as an important area for new discoveries within the next decade.  The MWA will be one of the first radio interferometers to attempt to detect redshifted 21 cm line emission from neutral hydrogen gas in the intergalactic medium (IGM) during this period.  The array has been designed to optimise its ability to detect brightness temperature fluctuations in the 21 cm line emission during the EoR in the redshift range $6 < z < 10$.    During the EoR, primordial neutral hydrogen begins to be ionised by the radiation from the first luminous sources.  The MWA has sufficient thermal sensitivity to detect the presence of the large ionised voids that form during reionisation through measurements of the power spectrum and other statistical properties of the fluctuations to a significance level of 14$\sigma$ \citep{bea12}.   In order to achieve this objective, the MWA will be a testbed to develop and demonstrate techniques to subtract the bright radio foregrounds that obscure the 21 cm background.

Radio emission from the Galaxy and from extragalactic sources is both a complicating foreground for EoR observations and an interesting scientific target that forms the second key science theme for the MWA.  The MWA will be unique in its ability to conduct an arc-minute resolution, confusion-limited survey of the full Southern Hemisphere sky below 10$^{\circ}$ declination over the 80$-$300 MHz frequency range.  The survey will include the Galactic Centre and the Large and Small Magellanic Clouds.   At the low observing frequencies of the MWA, non-thermal processes and Faraday rotation (and depolarization) effects will be prominent.  The MWA should be particularly well suited to identifying the missing population of old and faint supernova remnants (SNRs) in the Galaxy, closing the gap between the $\sim300$ known SNRs and the expected 1000 to 2000, and thereby providing a critical measurement of the total energy budget of the interstellar medium.  Additional experiments planned for the MWA target radio relics and clusters, the cosmic web, and Faraday tomography to probe magnetic fields.  Cosmic ray mapping may be also be possible along sight-lines where sufficiently dense HII regions have become optically thick in the MWA frequency band, blocking synchrotron emission from the Galaxy behind them.

With its high survey efficiency, as well as planned long integrations on EoR target fields, the MWA will enable sensitive transient and variable searches for both rare and faint events on timescales from seconds to days. The MWA will perform blind searches and target known transient sources, including low-mass stars and brown dwarfs, pulsars, X-ray binaries, and isolated neutron stars. At time resolution in the $\mu$s $-$ 1s range, the voltage capture capability of the MWA will allow studies of pulsars at low frequencies and searches for fast transients (e.g. \citet{way11}).

The final key science theme for the MWA encompasses the field of space weather, targeting multiple aspects of solar bursts as they travel from the surface of the Sun to the Earth.  The primary science focuses on high-dynamic range spectroscopic imaging to map the frequency, spatial, and time evolution of radio bursts occuring in the solar corona at heights of approximately 1 to 4 solar radii.  Interplanetary scintillation will be used to constrain the density and turbulence of the interstellar wind in the inner heliosphere.  The MWA may also enable measurements of the magnetic field of the heliosphere plasma if it proves possible to track changes in the polarisation angles of background sources.   Lastly, the calibration solutions of the MWA will yield near-real time corrections for ionospheric distortions, providing a new window into variability in the Earth's ionosphere.

\section{Discussion}

The MWA will be the first of the SKA Precursors to come to completion.  The MWA will also be the only low frequency SKA Precursor and the most capable general purpose low frequency radio telescope in the Southern Hemisphere, located on an extremely low interference site at the MRO.  The MWA is therefore a unique facility and highly complementary to ASKAP (also located at the MRO), with access to a similar field of view, but with ASKAP operating at higher frequencies (700 - 1800 MHz).  This frequency diversity potentially allows simultaneous coverage of interesting astronomical phenomena over the frequency range 80 - 1800 MHz from the MRO.  The MWA is also highly complementary to LOFAR, as the premier low frequency radio telesope in the Northern Hemisphere, providing complementary coverage of the whole sky and also longitudinal coverage important for observations of the Sun.

As one of a range of new facilities being brought online around the world now and in the future, many with an emphasis on high sensitivity and/or wide field of view survey science, a comparison of relevant metrics is useful.  In particular, a prime metric of importance for new survey science programs is the survey speed (SS), defined as \citep{cor07}:

\begin{equation}
SS = \frac{\Omega_{FoV}}{\tau},
\end{equation}

where $\Omega_{Fov}$ is the solid angle subtended by the field of view of the telescope in question, and $\tau$ is the time required for an image of that field of view to reach a particular level of sensitivity.  The wider the field of view of the primary array elements and the more primary elements in the array, the faster an array can survey a given region of the sky.  ASKAP, the MWA and LOFAR are well suited to survey science, as will the planned SKA.

\begin{figure*}[ht]
\begin{center}
\includegraphics[scale=0.5, angle=0]{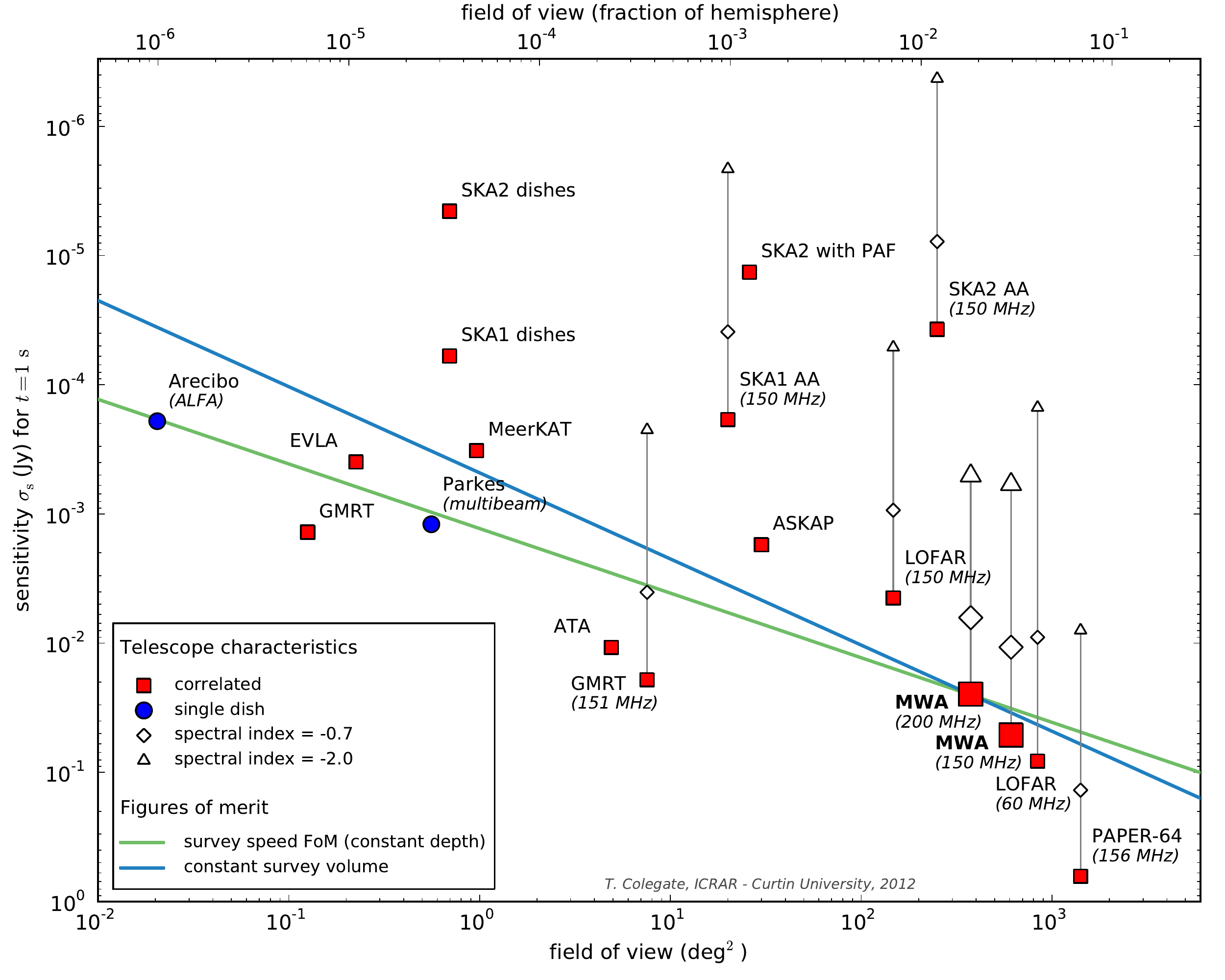}
\caption{Comparison of field of view and sensitivity for existing and future radio telescopes.  Red squares denote interferometric telescopes.  Blue circles denote single dish telescopes.  For the low frequency instruments, spectral index corrections to 1.4 GHz are indicated by the open diamonds and triangles, to allow comparison with higher frequency instruments.  Note that in this figure, sensitivity increases to the top of the figure ($\sigma_{s}$ decreases).  Figures of merit as discussed in the text are indicated by the blue and green lines.}\label{fig:surveyspeed}
\end{center}
\end{figure*}

A comparison of the survey speed metric is provided in Figure \ref{fig:surveyspeed}, for various existing and future instruments.

Figure \ref{fig:surveyspeed} shows that the MWA is highly competitive in terms of field of view on the sky, at an order of magnitude lower sensitivity than LOFAR, but of comparable sensitivity to the GMRT.  In producing this comparison, the low frequency instruments are shown with corrections for spectral indices of $-0.7$ and $-2.0$, corresponding to incoherent and coherent emission processes, respectively, in order to compare to the higher frequency instruments, which are all shown for an observing frequency of 1.4 GHz.  This style of comparison follows \citet{fen11}.  Additionally, in Figure \ref{fig:surveyspeed}, the aperture array instruments are assumed to be pointed at zenith and a common sky temperature is assumed at low frequencies following \citet{ska113}.  The solid lines shown denote constant depth and constant volume figures of merit, arbitrarily normalised to the MWA point at 200 MHz.

The MWA is therefore very well suited to large scale and repeated surveys of the sky between 80 and 300 MHz, down to the expected confusion limit of 10 - 20 mJy.  An expansion of the MWA at some future date to 256 tiles would move the MWA points in Figure \ref{fig:surveyspeed} higher by a factor of two in sensitivity.



In its formal role as an SKA Precursor, the MWA feeds information to the international SKA project, primarily by informing activities such as the development of conceptual design reviews for the low frequency component of the SKA, and by reporting technical and scientific results into international SKA fora.  A significant fraction of the MWA is built upon novel and new approaches to radio astronomy.  For example, the use of aperture arrays to implement steerable antennas with no moving parts is a pathfinder activity for the SKA and the MWA and LOFAR are pioneering this technology on medium to large scales.  The choice of siting the MWA at the MRO, as part of a greenfield development in search of the best environments for low frequency radio astronomy, clearly has massive benefits for low frequency science, but comes with significant challenges.  The practical challenges attendant to the construction of highly complex systems, in locations well away from the traditional centres of support for radio astronomy, are challenges that the SKA will have to face on an even larger scale.  Lessons learned from the MWA have a deep relevance to how the SKA project could proceed.

The MWA is expected to have an approximate five year lifetime, in order to acheive its science goals.  Beyond the operational lifetime of the facility, we note that the technologies that have gone into the MWA analog, digital and signal processing systems will be of order a decade old and it is likely that new, more efficient, more powerful, and less power intensive hardware will be available for the same task.  It is possible, therefore, to imagine a future re-use of the MWA infrastructure (power and fibre reticulation, capacity in CPF, and data transport to the Pawsey Centre) that could support a new and much more capable successor to the MWA, on a 5 - 10 year timescale.  This scenario highlights the enormous value created in building highly capable infrastructure at a remote and, from an RFI point of view, pristine site.  Such a scenario is supported by the fact that the fielded MWA equipment in tiles, beamformers and receivers, are all highly portable and can be easily decommissioned and replaced.  Thus, the MWA consortium recognises a long term future for low frequency radio astronomy at the MRO, through the MWA initially and then, perhaps, subsequent generations of instrument that re-use the underlying infrastructure.

An immediate possibility that takes advantage of the MWA infrastructure is for the facility to host some of the required prototyping activities for the low frequency sparse aperture array component of the SKA.  As the MRO is the internationally endorsed host site for the low frequency SKA in Phases 1 and 2, immediate advantage to the SKA project could be derived through use of the MWA infrastructure in testing and verifying the performance of prototype SKA hardware, through the international SKA Aperture Array Verification Programme (AAVP).

\section*{Acknowledgments}
We acknowledge the Wajarri Yamatji people as the traditional owners of the Observatory site.  Support for the MWA comes from the U.S. National Science Foundation (grants AST CAREER-0847753, AST-0457585, AST-0908884 and PHY-0835713), the Australian Research Council (LIEF grants LE0775621 and LE0882938), the U.S. Air Force Office of Scientic Research (grant FA9550-0510247), the Centre for All-sky Astrophysics (an Australian Research Council Centre of Excellence funded by grant CE110001020), New Zealand Ministry of Economic Development (grant MED-E1799), an IBM
Shared University Research Grant (via VUW \& Curtin), the Smithsonian Astrophysical Observatory, the MIT School of Science, the Raman Research Institute, the Australian National University, the Victoria University of
Wellington, the Australian Federal government via the National Collaborative Research Infrastructure Strategy, Education Investment Fund and the Australia India Strategic Research Fund and Astronomy Australia Limited, under contract to Curtin University, the iVEC Petabyte Data Store, the Initiative in Innovative Computing and NVIDIA sponsored CUDA Center for Excellence at Harvard, and the International Centre for Radio Astronomy Research, a Joint Venture of Curtin University and The University of Western Australia, funded by the Western Australian State government.



\begin{thebibliography}{}

\bibitem[Beardsley et al.(2012)]{bea12}Beardsley, A. et al. 2012, ApJ, submitted


\bibitem[Bell et al.(2012)]{bel12}Bell, M. et al. 2012, in preparation

\bibitem[Bernardi et al.(2012)]{Bernardi-etal.2012}Bernardi, G. et al. 2012, in preparation

\bibitem[Bernardi et al.(2011)]{Bernardi-etal.2011}Bernardi, G. et al. 2011, MNRAS, 413, 411

\bibitem[Bowman et al.(2012)]{bow12}Bowman, J. et al. 2012, in preparation

\bibitem[Bowman \& Rogers(2010)]{bow10}Bowman, J.D \& Rogers, A.E.E. 2010, Nature, 468, 796

\bibitem[Bracewell(1984)]{Bracewell1984}Bracewell, R.N. 1984, In J. A. Roberts, ed. {\it Indirect Imaging} Cambridge: Cambridge University Press, vol. 177

\bibitem[Bhatnagar et al.(2008)]{Bhatnagar-etal.2008}Bhatnagar, S., Cornwell, T.J., Golap, K. \& Uson, J.M. 2008, {\it A\&A}, 487(1), 419



\bibitem[Carilli \& Rawlings(2004)]{skascience}eds: Carilli, C. \& Rawlings, S. 2012, New Astronomy Rewiews, Vol. 48, Elsevier

\bibitem[Clark et al.(2012)]{Clarketal}Clark, La Plante \& Greenhill 2012, arXiv 1107.4264

\bibitem[Cornwell(2005)]{Cornwell2005}Cornwell, T.J., Golap, K. \& Bhatnagar, S. 2005, in {\it ASP Conf. Series: Astronomical Data Analysis Software and Systems XIV}, San Francisco: Astronomical Society of the Pacific, vol. 347, pp. 86

\bibitem[Cornwell \& Perley(1992)]{CornwellPerley1992}Cornwell, T.J. \& Perley, R.A. 1992, {\it A\&A}, 261, 353

\bibitem[Cordes(2010)]{cor07}Cordes, J.M. 2007, SKA Memo \#109, "Survey metrics", www.skatelescope.org/pages/page\_memos.htm

\bibitem[Deller et al.(2011)]{del07}Deller, A.T. et al. 2011, PASP, 123, 275

\bibitem[Deller et al.(2007)]{del11}Deller, A.T., Tingay, S.J., Bailes, M. \& West, C. 2007, PASP, 119, 318

\bibitem[Dewdney et al.(2010)]{ska}Dewdney, P.. et al. 2010, SKA Memo \#130, "SKA Phase 1: Prelimninary System Description", www.skatelescope.org/pages/page\_memos.htm

\bibitem[Edgar et al.(2010)]{Edgar-etal.2010}Edgar, R.G. et al. 2010, {\it Comp. Phys. Comm.}, 181(10), 1707

\bibitem[Fender \& Bell(2011)]{fen11} Fender, R. \& Bell, M.E. 2011, BASI, 39, 315

\bibitem[Gorski et al.(2005)]{Gorski-etal.2005}Gorski, K.M. et al. 2005, {\it ApJ}, 622(2), 759

\bibitem[Hamaker(2000)]{Hamaker2000}Hamaker, J.P. 2000, {\it Astron. Astrophys. Suppl. Ser.}, vol. 143, pp. 515

\bibitem[Hazelton et al.(2012)]{haz12}Hazelton, B. et al. 2012, in preparation

\bibitem[Johnston et al.(2008)]{jon08}Johnston, S. et al. 2008, Ex.A., 22, 151

\bibitem[Johnston et al.(2007)]{jon07}Johnston, S. et al. 2007, PASA, 24, 174

\bibitem[Lonsdale et al.(2009)]{Lonsdale-etal.2009}Lonsdale, C.J. et al.2009, Proc. IEEE, 97, 1497

\bibitem[McKinley et al.(2012)]{mck12}McKinley, B. et al. 2012, in preparation

\bibitem[Mitchell et al.(2012)]{Mitchell-etal.2012}Mitchell, D.A., Wayth, R.B., Bernardi, G., Greenhill, L.J. \& Ord, S.M. 2012, submitted

\bibitem[Mitchell et al.(2010)]{Mitchell-etal.2010}Mitchell, D.A. et al. 2010, PoS(RFI2010)016

\bibitem[Mitchell et al.(2008)]{Mitchell-etal.2008}Mitchell, D.A. et al. 2008, {\it IEEE Journal of Selected Topics in Signal Processing}, 2, 1993

\bibitem[Morales \& Matejek(2009)]{MoralesMatejek2009}Morales, M.F. \& Matejek, M. 2009, {\it MNRAS}, 400, 1814

\bibitem[Myers et al.(2003)]{Myers-etal.2003}Myers, S.T. et al. 2003, {\it ApJ}, 591, 575

\bibitem[Nieuwpoort \& Romein(2009)]{NieuwpoortRomein}Nieuwpoort, R.V.V. \& Romein, J.W. 2009, In 23rd ACM International Conference on Supercomputing

\bibitem[Nijboer, Pandey-Pommier \& de Bruyn(2009)]{ska113}Nijboer, R.J. Pandey-Pommier, M. \& de Bruyn, A.G. 2009, SKA Memo \#113, "LOFAR imaging capabilities and system sensitivity", www.skatelescope.org/pages/page\_memos.htm

\bibitem[Nordam(2004)]{Noordam2004}Noordam, J.E. 2004, in {\it  Proc. SPIE: Groundbased Telescopes}, Glasgow, vol. 5489, pp. 817

\bibitem[Oberoi et al.(2011)]{obe11}Oberoi, D. et al. 2011, ApJ, 728, L27

\bibitem[Ord et al.(2010)]{Ord-etal.2010}Ord, S,M. et al. 2010, {\it PASP}, 122, 1353

\bibitem[Ord et al.(2009)]{Ord-etal.2009}Ord, S.M. et al. 2009, {\it Proc. of ADASS XVII}

\bibitem[Perley(1999)]{Perley1999}Perley, R.A. 1999, In G. B. Taylor, C. L. Carilli, and R. A. Perley (editors), {\it Synthesis Imaging in Radio Astronomy II}, ASP Conf. Series, vol. 180, pp. 383

\bibitem[Pindor et al.(2011)]{Pindor-etal.2011}Pindor, B., Wyithe, J.S.B., Mitchell, D.A., Ord, S.M., Wayth, R.B. \& Greenhill, L.J. 2011, {\it PASA}, 28 (1), 46


\bibitem[Smirnov(2011)]{Smirnov2011}Smirnov, O.M. 2011, {\it A\&A}, 527

\bibitem[Sullivan et al.(2012)]{Sullivan-etal.2012}Sullivan, I.S. et al. 2012, {\it In prep.}

\bibitem[Thompson et al.(2001)]{Thompson-etal.2001}Thompson, A.R., Moran, J.M. \& Swenson Jr., G.W. 2001, {\it Interferometry and Synthesis in Radio Astronomy}, 2nd Ed., New York: Wiley

\bibitem[van Harlem et al.(2012)]{lofar}van Harlem, M. et al. 2012, in preparation

\bibitem[Wayth et al.(2012)]{way11}Wayth R.B. et al. 2012, ApJ accepted (arXiv:1205.5840)

\bibitem[Wells, Greisen \& Harten(1981)]{Wellsetal1981}Wells, Greisen \& Harten 1981, A\&A Supplement, Vol. 44, P. 363

\bibitem[Wicenec \& Knudstrup(2007)]{wic07}Wicenec, A. \& Knudstrup, J. 2007, The Messenger, 129, 27

\bibitem[Williams et al.(2012)]{Williams-etal.2012}Williams, C.L. et al. 2012, ApJ accepted (arXiv:1203.5790)

\end{thebibliography}
\end{document}